\documentstyle[aps]{revtex}              
\begin{document}
\draft
\twocolumn [\hsize\textwidth\columnwidth\hsize\csname  
@twocolumnfalse\endcsname              
\title{\bf Dynamics of the reaction-diffusion
system $A + B \rightarrow 0 $ with input of particles}
\author{ Boris ~M.~Shipilevsky}
\address{ Institute of Solid State Physics, Chernogolovka,
Moscow district, 142432, Russia}
\date{\today}
\maketitle
\begin{abstract}

We study dynamics of filling of an initially empty finite medium by diffusing
particles $A$ and $B$, which arise on the surface upon dissociation of $AB$
molecules, impinging on it with a fixed flux density $I$, and desorb from it
by the reaction $A + B\rightarrow AB\rightarrow 0$. We show that once the
bulk diffusivities differ ($p=D_{A}/D_{B}<1$), there exists a critical flux
density $I_{c}(p)$, above which the relaxation dynamics to the steady state
is qualitatively changed: on time dependences of $c_{As}/c_{e}$ ($c_{e}$ being
the steady state concentration at $t\rightarrow \infty$) a maximum appears,
the amplitude of which grows both with $I$ and with $D_{B}/D_{A}$ ratio.
In the diffusion-controlled limit $I \gg I_{c}$ at $p \ll 1$ the reaction
"selects" the {\it universal laws} for the particles number growth
${\cal N}_{A}={\cal N}_{B}\propto t^{1/4}$ and the evolution of the surface
concentrations $c_{As}\propto t^{-1/4},c_{Bs}\propto t^{1/4}$, which are
approached by one of the {\it two characteristic regimes} i) and ii) with
the corresponding hierarhy of the intermediate power-law asymptotics.
In the first of these $c_{As}$ goes through a comparatively {\it sharp}
max$(c_{As}/c_{e})\propto I^{1/6}$, the amplitude of which is
$p$-independent, in the second one $c_{As}$ goes through a {\it plateau-like}
max$(c_{As}/c_{e})\propto p^{-1/4}$, the amplitude of which is
$I$-independent. We demonstrate that on the main filling stage the evolution
of the ${\cal N}(t)/{\cal N}_{e}, c_{As}(t)/c_{e},$ and $c_{Bs}(t)/c_{e}$
trajectories with changing $p$ or $J$ between the limiting regimes i) and ii)
is unambigouosly defined by the value of the scaling parameter
${\cal K}=p^{3/2}J$ ($J$ being the reduced flux density) and is described by
the set of {\it scaling laws}, which we study in detail analytically and
numerically. In conclusion, we analyze specific features of the long-time
relaxation dynamics and calculate exactly the relaxation rate $\omega(p,J)$.

\end{abstract}
\pacs{82.20Mj, 05.70 Ln}
]                                  

\narrowtext

\section{INTRODUCTION}

 For the last decade the reaction-diffusion system $A + B \rightarrow 0$,
where unlike species $A$ and $B$ diffuse and irreversibly react in the bulk
of a $d$-dimensional substrate, has acquired the status of one of the most
popular objects in nonequilibrium statistical physics \cite{rev}.
Two situations have been investigated most intensively: (i) dynamics of the
$A +B \rightarrow 0$ annihilation in an infinite system with initially
homogeneously (randomly) and equimolarly distributed reactants, in which
below the critical dimension $d_{c}=4$ a dynamical clustering develops
(fluctuation-induced like-particle domains formation) and, as a result,
an anomalous reaction deceleration arises (Ovchinnikov and Zeldovich,
1978 \cite{ov}, Toussaint and Wilczek, 1983 \cite{to}); (ii) behaviour of the
dynamic reaction front in an infinite system with initially spatially
separated reactants (Galfi and Racz, 1988 \cite{ga}), and structure of the
steady state front in a finite system, at the ends of which are injected
equal currents of $A's$ and $B's$ particles (Ben-Naim and Redner, 1992
\cite{ben}, Cornell and Droz, 1993 \cite{cor}).

Recently, we have shown \cite{s4,s7}, that in another wide class of RD
system, where reaction and diffusion are spatially separated ( i.e. reaction
proceeds on the surface of the medium and diffusion proceeds in its bulk )
the interplay between reaction and diffusion acquires qualitatively new
features and leads to the threshold self-organizing dynamics of the
$A+B\rightarrow 0$. It has been found that once particles $A$ and $B$ diffuse
at different mobilities from the bulk of finite medium onto the surface and
die on it by the reaction $A + B \rightarrow 0$, there should exist some
threshold difference in the initial numbers of $A$ and $B$ particles ,
$\Delta_{c}$, above which the process of their death , instead of usual
deceleration, starts {\it to accelerate autocatalytically}. Moreover, it has
been demonstrated \cite{s9} that in the diffusion-controlled limit $\Delta
\rightarrow \infty$ a new critical phenomenon develops in the system - {\it
annihilation catastrophe}, which arises as a result of self-organizing
explosive growth (drop) of the surface concentrations of, respectively,
slow and fast particles ({\it concentration explosion}) and manifests itself
in the form of an abrupt singular jump in the desorption flux relaxation rate.
In the limit of strong difference of diffusivities the annihilation
catastrophe leads to the phenomenon of abrupt disappearance of the flux
({\it flux breaking effect}), which may pretend to be one of the most
dramatic manifestations of the reaction-diffusion interplay.

The aim of this paper is, in contrast to the pure annihilation
problem $A + B \rightarrow 0$, to consider dynamics of filling of an
{\it initially empty} finite medium by diffusing particles $A$ and $B$,
which arise on the surface upon dissociation of $AB$ molecules, impinging on
it with a {\it fixed} flux density $I$, and desorb from it by the reaction
$A + B \rightarrow AB \rightarrow 0$. We assume that the density of the
input flux $I$ is not too large and (or) the reaction rate constant
is sufficiently large so that in the process of filling up to the steady
state the surface coverages by adatoms $A_{ads}$ and $B_{ads}$ remain
small enough and, therefore, the reflection of $AB$ molecules from the
occupied sites can be neglected. In addition, we assume that the particles
exchange rates between the surface and the subsurface layer are great, and
already at early stages of diffusion into the bulk a quasiequilibrium is
reached between the surface and the subsurface layer. Finally, we assume
that in our $2d$-reaction + $3d$-diffusion problem diffusion smoothes out
lateral fluctuations and, hence, planar distribution of $A's$ and $B's$
particles is sustained mesoscopically uniform, so that 1) the desorption flux
density may be described by the mean-field expression and 2) the problem can
effectively be considered as one dimensional. In the framework of these
assumptions we formulate a closed system of nonlinear boundary value
diffusion problems, which we then investigate in detail analytically and
numerically.

\section{MODEL}

Let the both surfaces $X=\pm\ell$ of an initially empty infinitely
extended slab of thickness $2\ell$ be hit with a fixed-density $I$ flux
of $AB$ molecules, which dissociate on unoccupied sites to adatoms $A_{ads}$
and $B_{ads}$ with probability 1. Adatoms $A_{ads}$ and $B_{ads}$ migrate
along the surface and either, on running into one another, irreversibly
desorb by the reaction $A_{ads}+B_{ads}\rightarrow AB\rightarrow 0$ or
diffuse into the bulk, gradually filling it to some steady state
concentration $c_{e}(I)$ ( Fig.1 ).
Because of planar spatial homogeneity (see below), the bulk diffusion flux
must be directed normally to the surface plane, that is, the problem is
effectiveliy one dimensional. The boundary conditions for the bulk diffusion
equations

\begin{eqnarray}
\partial_{t}c_{i}=D_{i}\partial^{2}_{X}c_{i}
\end{eqnarray}

($i=A,B$) can be derived from the balance of the flux densities on
the surface

\begin{eqnarray}
\dot {\rho_{i}}=I^{a}-I^{d}-I_{i}^{sb}
\end{eqnarray}

and in the subsurface layer

\begin{eqnarray}
a\dot c_{is}=I_{i}^{sb}-I^{D}_{i}\mid_{s}.
\end{eqnarray}

Here $\rho_{i}$ is the surface concentration of $i$-adatoms
($cm^{-2}$), $c_{is}= c_{i}\mid_{X=\ell}$ is the concentration of
$i$-particles in the subsurface layer ($cm^{-3}$), $a$ is the lattice
parameter, $I^{a}$ and $I^{d}$ are adsorption and desorption flux
densities,
$I^{sb}_{i} = {\cal I}^{s}_{i} - {\cal I}^{b}_{i} =
\Gamma^{s}_{i}\rho_{i}-\Gamma^{b}_{i}c_{is}\theta_{0}$
is the surface-to-subsurface layer flux density, where $\Gamma^{s}_{i}$
and $\Gamma^{b}_{i}$ are the rate constants of $i$- particles transition
from the surface into the subsurface layer and back, respectively,
$\theta_{0}=1-\theta_{A}-\theta_{B}$ is the fraction of vacant sites on
the surface, which we take to be close to unity ($\theta_{i} \ll 1$),
$I^{D}_{i}\mid_{s}=D_{i}\partial_{X}c_{i}\mid_{X=\ell}$ is the diffusion
flux density at the surface (by symmetry, we consider the interval
$[0,\ell]$ only with the condition $\partial_{X}c_{i}\mid_{X=0}=0$).
We assume that the subsurface layer - surface barrier is not much
different from the diffusion barrier in the bulk
($\Gamma_{i}^{b}\sim D_{i}/a$), and probability of the adatoms transition
to the subsurface layer is much greater than probability of their
desorption (${\cal I}^{s}_{i} \gg I^{d}$). Then, prior to
desorbing, the particles multiply go to the subsurface layer and back,
and already at early stages of diffusion into the bulk a quasiequilibrium
must be established between the surface and the subsurface layer,
$\rho_{i}\simeq f_{i}c_{is}$, where $f_{i}=\Gamma^{b}_{i}/\Gamma^{s}_{i}$
is the surface segregation coefficient. Moreover, if the surface
segregation is not enough strong (at elevated temperatures usually
$f\sim (1-10^{2})a$), and the system size is quite large,
$\ell \gg f_{i}$, then at comparatively short (in $\ell^{2}/D_{i}$ scale)
times $f_{i} \ll \ell^{D}_{i}=\sqrt {D_{i}t} \ll \ell$, when the
number of particles in the bulk (per surface unit)
${\cal N}_{i}=\int_{0}^{\ell}{c_{i}dX} \sim c_{is}\sqrt {D_{i}t}$
much exceeds that on the surface $f_{i}c_{is}$, the capacities of the
subsurface layer ($I^{D}_{i}\mid_{s}=I_{i}^{sb}$) and
of the surface ($f_{i}\dot {|c_{is}|}\ll I^{D}_{i}\mid_{s}$)
may be neglected, therefore conditions (2) and (3) are degenerated to the
following

\begin{eqnarray}
\nonumber
\dot {\cal N}_{i}= I^{D}_{i}\mid_{s} = I^{a}-I^{d}.
\end{eqnarray}

According to Refs. \cite{lin}, \cite{do}, and \cite{cle}, in a distributed 
$d$- dimensional system $A+B\rightarrow0$ with injection of $A$ and $B$ 
particles, with their difference being strictly conserved 
(correlated landing), a critical (marginal) dimensionality, above which no 
fluctuation-induced segregation of $A$ and $B$ occurs, equals 
$d_{c}=2$ ($2d$-reaction + $2d$-diffusion).
It is to be expected that in our case ($2d$-reaction + $3d$-diffusion)
$3d$-diffusion smoothes out lateral fluctuations, and a planar distribution
of $A$ and $B$ particles is sustained mesoscopically uniform. So, the
desorption flux density may be described by the mean-field expression
$I^{d}=k\rho_{A}\rho_{B}$. From the condition $I^{sb}_{i}=I^{a}-I^{d}$
it follows

$$
\rho_{i}= f_{i}c_{is}\theta_{0}[1+(I^{a}-I^{d})/{\cal I}^{b}_{i}],
$$

therefore, taking $I^{a}-I^{d} \ll {\cal I}^{b}_{i}$, we obtain

$$
I^{d} = \kappa c_{As}c_{Bs}\theta^{2}_{0}
[1+(I^{a}-I^{d})(1/{\cal I}^{b}_{A}+1/{\cal I}^{b}_{B})],
$$

where $\kappa=k f_{A}f_{B}$ is the effective reaction rate
constant ($cm^{4}/s$). The above requirement
$I^{a},I^{d}\ll {\cal I}^{s}_{i}\simeq {\cal I}^{b}_{i}$
imposes on the concentrations limitations
$I^{a}/\Gamma_{A}^{b}\ll c_{As} \ll \Gamma_{B}^{b}/\kappa,
I^{a}/\Gamma_{B}^{b} \ll c_{Bs} \ll \Gamma_{A}^{b}/\kappa$, the lower
boundary of which defines the conditions for the crossover to the
subsurface layer-surface quasiequilibrium regime. Restrictions on the
flux densities $I^{d},I^{a}\ll \Gamma_{A}^{b}\Gamma_{B}^{b}/\kappa$
follow herefrom. Taking into account the reflection from occupied sites,
we write $I^{a}=I\theta_{0}^{2}$ to obtain
$I^{a}-I^{d}=\chi\theta_{0}^{2}(I-\kappa c_{As}c_{Bs})$, where
factor $\chi=  [1+\theta_{0}((\kappa/\Gamma_{B}^{b})c_{As}+
(\kappa/\Gamma^{b}_{A})c_{Bs})]^{-1}\simeq 1$. So, after a short-term
transient stage $t \gg t_{tr}=$ max$(f_{i}^{2}/D_{i})$
the boundary conditions take the form

\begin{eqnarray}
\dot{\cal N}_{i}=D_{i}\partial_{X}c_{i}\mid_{s}=\chi\theta_{0}^{2}
(I -\kappa c_{As}c_{Bs}),
\end{eqnarray}

where $1-\chi\theta_{0}^{2} \ll 1$. We assume that on a transient stage the
desorption can be neglected, i.e. $t_{tr}\ll t_{q}$, where $t_{q}$ is a
characteristic time at which the desorption flux becomes comparable with
the input flux $I^{d}\sim I$. Using then the Laplace transform, at
$t \ll t_{q},t_{i}^{D}=\ell^{2}/D_{i}$ one can easily obtain the complete
solution of (1)-(3), wherefrom at $t \gg t_{if} \gg t_{is}$
(here $t_{is}= D_{i}/(\Gamma_{i}^{b})^{2}$ and $t_{if}=f_{i}^{2}/D_{i}$)
it follows

\begin{eqnarray}
\nonumber
{\cal N}_{i}={\cal N}_{i}^{(0)}\left[1 - \frac{2}{\sqrt {\pi}}\left(\frac
{t_{if}}{t}\right)^{1/2} + \cdots\right],
\end{eqnarray}
\begin{eqnarray}
c_{is}=c_{is}^{(0)}\left[1 -\frac {\sqrt {\pi}}{2}\left(\frac
{t_{if}}{t}\right)^{1/2}+ \cdots\right],
\end{eqnarray}
\begin{eqnarray}
\nonumber
\rho_{i}=f_{i}c_{is}\left[1 + \frac {\sqrt {\pi}}{2}\left(\frac
{t_{is}}{t}\right)^{1/2} + \cdots\right],
\end{eqnarray}

where ${\cal N}_{i}^{(0)}=It$ and $c_{is}^{(0)}=\frac {2I\sqrt {t}}
{\sqrt {\pi D_{i}}}$ are the solutions of (1),(4) with the initial conditions
$c_{i}(X,0)=0$ at $\chi \theta_{0}^{2}=1$ and $I^{d}/I \rightarrow 0$.
From (5) it is seen that the influence of the transient stage rapidly decays
in time. So, as we are mainly interested here in the system's behavior at
$t \gg t_{tr}$, we shall take $c_{i}(X,0)=0$ as the initial condition for
(1), (4). According to (5), the condition
$I^{d}\sim \kappa I^{2}t/\sqrt {D_{A}D_{B}}\ll I$
is reduced to the requirement
$t \ll t_{q}= \frac {\sqrt {D_{A}D_{B}}}{I\kappa}$,
whence it follows $I \ll \sqrt {D_{A}D_{B}}/\kappa t_{tr}$.
Introducing the index $"H"$ ({\it heavy}) for the slower diffusing species
and the index $"L"$ ({\it light}) for the faster one, taking
$\chi\theta_{0}^{2}=1$ and going to dimensionless variables, we come
finally to the boundary value diffusion problem

\begin{mathletters}
\label{ao}
\begin{eqnarray}
\partial h/\partial \tau  = \nabla^{2}h \quad , \quad
\partial l/\partial \tau  = (1/p)\nabla^{2}l,
\end{eqnarray}
\begin{eqnarray}
\label{bo}
\nabla h \mid_{s} = (1/p)\nabla l\mid_{s}
=J - h_{s}l_{s},
\end{eqnarray}
\end{mathletters}

with the conditions of symmetry $\nabla (h,l) \mid_{x=0}=0$ and the initial
conditions $h(x,0)=l(x,0)=0$. Here $h(x,\tau)=c_{H}/c_{*}$ and
$l(x,\tau) = c_{L}/c_{*}$ are the reduced concentrations, $J=I/I_{*}$ is
the reduced flux density, $\nabla=\equiv \partial/\partial x$
$x = X/\ell \in [0,1]$ is the nondimensional coordinate,
$\tau = D_{H}t/\ell^{2}$ is the nondimensional time,
$p = D_{H}/D_{L} \le  1$ is the ratio of species diffusivities,
$I_{*}=\kappa c_{*}^{2}=(D_{H}/\sqrt {k}\ell)^{2}$
and $c_{*} = D_{H}/\kappa \ell$ are the characteristic flux density and
concentration scales, from which the relaxation dynamics to the steady-state
becomes diffusion-controlled (in the reaction-controlled regime the both
species are distributed uniformly $c_{is}={\cal N}/\ell = It/\ell$, so,
from the condition $I^{d}\sim \kappa(It/\ell)^{2}\ll I$ it follows
$t_{R}\sim \ell/\sqrt{\kappa I}$. By comparing $t_{R}$ with the
characteristic diffusion time of heavy species
$t_{i}^{D}\sim \ell^{2}/D_{H}$, we obtain
$I_{*}=(D_{H}/\sqrt{\kappa}\ell)^{2}$).

According to (5), with an accuracy to a neglibly small capacity of the
surface layer ($\tau \gg \tau_{tr}=$ max $(d^{2}_{H}, p d^{2}_{L})$,
$d_{i}=f_{i}/\ell\ll 1$), equal amounts (per surface unit) of $H$ and $L$
particles diffuse into the bulk ${\cal N}_{H}={\cal N}_{L}={\cal N}$,
therefore in the steady state $(\tau \rightarrow \infty)$

\begin{eqnarray}
\nonumber
h_{e}=l_{e}=N_{e}=\sqrt {J},
\end{eqnarray}

where $N={\cal N}/{\cal N}_{*}$ is the reduced number of particles and
${\cal N}_{*}=c_{*}\ell=D_{H}/\kappa$ is the characteristic scale of
the number of particles. In dimensionless variables requirement
$1-\chi \ll 1$ with account of $\Gamma_{i}^{b}\sim D_{i}/a$ leads to the
conditions $h_{s} \ll \Gamma_{L}^{b}\ell/D_{H} \sim p^{-1}(\ell/a)$
and $l_{s} \ll \Gamma_{H}^{b}\ell/D_{H} \sim \ell/a $ whence there follows
a limitation on the reduced flux density $l_{e}=h_{e}=\sqrt {J} \ll \ell/a$
and, therefore, $J \ll J_{\chi}^{u} \sim (\ell/a)^{2}$. The requirement
$\theta_{i}\sim a^{2}\rho_{i}\ll 1$ leads to the conditions
$h_{s} \ll (\kappa/a^{2}D_{H})d_{H}^{-1}$ and
$l_{s}\ll (\kappa/a^{2}D_{H})d_{L}^{-1}$, whence by assuming
$d_{H}\sim d_{L}$ we obtain
$J \ll J_{\theta}^{u} \sim (\kappa/a^{2}D_{H})^{2}d_{H}^{-2}$.
Finally, the requirement $\tau_{tr} \ll \tau_{q}= 1/J\sqrt {p}$ leads to the
condition $J \ll J_{tr}^{u} \sim d_{H}^{-2}/\sqrt {p}$. In macroscopic
systems the quantities $J^{u}_{\chi}$ and $J^{u}_{tr}$ are very large
(for example at $\ell\simeq 10^{-1}cm$ and $f_{H}\sim f_{L}\sim 10a$
we have $J^{u}_{\chi}\sim 10^{14}$ and $J^{u}_{tr}\sim 10^{12}/\sqrt{p}$).
At temperatures of intensive desorption ($\kappa/a^{2}D_{H} \geq 1$) the
quantity $J_{\theta}^{u}$ is as large. So, the reduced flux density $J$
can, formally, be taken as being unlimitedly variable.

\section{TRANSITION IN RELAXATION DYNAMICS}

At a high enough density of the input flux, $J$, a quasiequilibrium
$J^{d}=h_{s}l_{s}\simeq J$ should be established well before the
particles distribution in the bulk becomes uniform. In view of the fact
that at $p<1$ there must be $l_{s}<h_{s}$, this means $h_{s}/h_{e}>1$, i.e.
at sufficiently large $J$ the $H$ particles surface concentration should
initially grow to some maximum $h_{s}^{M}>h_{e}$ and then, as a result of
diffusion into the bulk, relax asymtotically to its steady state value,
$h_{e}$, from above. We thus conclude that at $p < 1$ there ought to
exist a critical flux density $J_{c}(p)$, above which the relaxation
character is qualitatively changed. In this section we give a {\it linear}
analysis of the long-time relaxation dynamics, in terms of which we find the
relaxation rate $\omega(J,p)$ and the exact value of $J_{c}(p)$.

We introduce the new variables $\tilde{h}=h_{e}-h, \tilde{l}=l_{e}-l$ and
$\tilde{n}=N_{e}-N$. Then, instead of (6) we have

\begin{mathletters}
\label{ao}
\begin{eqnarray}
\partial \tilde{h}/\partial \tau =\nabla^{2}\tilde{h}, \quad
\partial \tilde{l}/\partial \tau =(1/p)\nabla^{2}\tilde{l},
\end{eqnarray}
\begin{eqnarray}
\label{bo}
\nabla \tilde{h}\mid_{s}=(1/p)\nabla \tilde {l}\mid_{s}=
-\sqrt {J}(\tilde {h}_{s} +\tilde{l}_{s})+\tilde {h}_{s}
\tilde {l}_{s}
\end{eqnarray}
\end{mathletters}

with $\nabla(\tilde {h},\tilde {l})\mid_{x=0}=0$. In a long-time limit
$\tilde {h},\tilde {l}\mid_{\tau\rightarrow \infty}\rightarrow 0$
the nonlinear term may be neglected, so by satisfying (7), we find
the leading terms of the long-time relaxation in the form

\begin{eqnarray}
\tilde {h}={\cal A}_{H}\cos(\sqrt {\omega}x)e^{-\omega\tau}, \quad
\tilde {l}={\cal A}_{L}\cos(\sqrt {p\omega}x)e^{-\omega\tau},
\end{eqnarray}

where ${\cal A}_{L}=(\sqrt {p}\sin\sqrt {\omega}/\sin\sqrt {p\omega})
{\cal A}_{H}$, and the relaxation rate $\omega(p,J)$ is defined by the least
positive root of the equation

\begin{eqnarray}
\cot \sqrt {\omega} =\sqrt {\frac {\omega}{J}}-\sqrt {p}\cot \sqrt {p\omega}
\end{eqnarray}

It follows from (8) that long-time asymptotics of surface concentrations
as a function of the particles number
$\tilde{n}=\int_{0}^{1}{\tilde {h}dx}=\int_{0}^{1}{\tilde {l}dx}$
has the form

\begin{eqnarray}
\tilde {h}_{s}\mid_{\tilde{n}\rightarrow 0}=
(\sqrt {\omega}\cot \sqrt {\omega})\tilde{n},
\tilde {l}_{s}\mid_{\tilde{n}\rightarrow 0}=
(\sqrt {p\omega}\cot \sqrt {p\omega})\tilde{n},
\end{eqnarray}

whence

\begin{eqnarray}
r_{s}=\frac {\tilde {h}_{s}}{\tilde {l}_{s}}\mid_{\tilde{n}\rightarrow 0}=
\frac {\cot \sqrt {\omega}}{\sqrt {p}\cot \sqrt {p\omega}}=
-1 + \sqrt {\frac {\omega}{Jp}}\tan(\sqrt {p\omega}).
\end{eqnarray}

According to (9), with growing $J$ the relaxation rate is increased from
$\omega=2 \sqrt {J}$ in the reaction-controlled limit $\sqrt {J} \ll 1$

\begin{eqnarray}
\nonumber
\omega= 2 \sqrt {J}(1- \frac {1+p}{3}\sqrt {J}+ \cdots ), \quad \sqrt {J} \ll 1
\end{eqnarray}

to a maximal

\begin{eqnarray}\nonumber
\omega=\omega_{m}(p)(1- b(p)/\sqrt {J}+ \cdots ), \quad \sqrt {J} \gg 1
\end{eqnarray}

in the diffusion-controlled limit $\sqrt {J} \gg 1$, the maximal relaxation
rate $\omega_{m}(p)$ growing herewith from

\begin{eqnarray}
\nonumber
\omega_{m}(1) = \pi^{2}/4
\end{eqnarray}

at $p\rightarrow 1$ to

\begin{eqnarray}
\nonumber
\omega_{m}(0)= 4.1158...
\end{eqnarray}

at $p\rightarrow 0$ ($b(p)$ changes from $b(1)=1$ to $b(0)=1.348...$).
So, in accord with (9),(10), we conclude that at $p< 1$ there exists
a critical relaxation rate $\omega_{c}=\pi^{2}/4$, which is reached at
a critical flux density

\begin{eqnarray}
J_{c}=\frac {\pi^{2}}{4p}\tan^{2}(\pi\sqrt {p}/2),
\end{eqnarray}

above which the quantity $\tilde {h}_{s}$ reverses its sign
$( + \rightarrow - )$, i.e. the character of the surface concentration
relaxation of $H$ particles qualitatively changes: at $J < J_{c}$ the
$h_{s}$ value grows to $h_{e}$ monotonously, whereas at $J > J_{c}$ the value
of $h_{s}$ reaches initially a maximum and then relaxes to $h_{e}$
from above (Fig.2). According to (11), as $J$ grows, the ratio
$r_{s}=\tilde h_{s}/\tilde l_{s}\mid_{n \rightarrow 0}$ is changed from $1$
at $\sqrt {J/J_{c}} \ll 1$

\begin{eqnarray}
\nonumber
r_{s}=1-\frac{2}{3} \sqrt {J}(1-p)+\cdots, \quad \sqrt {J/J_{c}} \ll 1
\end{eqnarray}

to  $-1$ at $\sqrt {J/J_{c}} \gg 1$

\begin{eqnarray}
\nonumber
r_{s}=-1 + \sqrt \frac {\omega_{m}}{Jp}\tan \sqrt {p\omega_{m}},
\quad \sqrt {J/J_{c}} \gg 1
\end{eqnarray}

as is illustrated for $p=0.1$ in Figs.3 and 4, the first of which
demonstrates the dependences of $r_{s}$ on $J/J_{c}$, calculated from
(9), (11), the second one demonstrates the time dependences of
$h_{s}/h_{e}$ and $l_{s}/l_{e}$ at $J=J_{c}$ and $J=10^{3}$, numerically
calculated from (6).

Eqs. (10) and (11) carry no information on the initial conditions,
so they, clearly, remain valid at any initial number of particles
${\cal N}_{H}(0)={\cal N}_{L}(0)={\cal N}(0)$ and at their arbitrary
initial distribution. In this case, according to (11), if the initial
distribution is uniform, and $\tilde{n}>0$, i.e. the initial particles number
is less than the steady state one, then at $J > J_{c}$ the $h_{s}$ value
passes a maximum and relaxes to the steady state level {\it from above},
whereas if the initial distribution is uniform, and $\tilde{n}<0$, i.e.
the initial particles number is greater than the steady state one, then at
$J > J_{c}$ the $h_{s}$ value passes a mimimum and relaxes to the steady
state level {\it from below}.

In the present paper the main attention will be focused on the behavior of
the surface concentrations and the growth dynamics of the particles number in
the diffusion-controlled limit $J \gg J_{c}$, which is of greatest interest.
As at $\omega\tau > 1$ deviations from the steady state become
exponentially small then, clearly, in the diffusion-controlled limit
the main nonlinear kinetic effects evolve in times $\tau \ll 1$
when the diffusion of $H$ particles proceeds, actually, into a semi-infinite
medium. In the next Section we shall consider the dynamics of behavior of
$h_{s}$,$l_{s}$ and $N$ at $\tau \ll 1$, taking into account the reflection
from the boundary $x=0$, and then, in Section V, we shall come back to an
analysis of the long-time relaxation and obtain expressions for the
${\cal A}_{i}(J,p)$ amplitudes.

\section{DIFFUSION OF H PARTICLES INTO A SEMI-INFINITE MEDIUM
$(\tau \ll 1)$.}

By applying the Laplace transform $\hat {f}(s)=\hat {\cal L}f(\tau)=
\int_{0}^{\infty} {e^{-s\tau}f(\tau)d\tau}$ to Eqs. (6), one can easily
obtain a formal solution of problem (6) in the form

\begin{eqnarray}
\hat {h}(x,s)=\hat {h}_{s}\frac{\cosh(x\sqrt {s})}{\cosh\sqrt {s}}, \quad
\hat {l}(x,s)=\hat {l}_{s}\frac{\cosh(x\sqrt {sp})}{\cosh\sqrt {sp}}
\end{eqnarray}

with the boundary conditions (6b), which can, for convenience, be represented
as

\begin{eqnarray}
\hat {N}= \frac {\hat {h}_{s}}{\sqrt {s}}\tanh\sqrt {s}=
\frac {\hat {l}_{s}}{\sqrt {sp}}\tanh\sqrt {sp}
\end{eqnarray}

and

\begin{eqnarray}
\hat {N}= \frac {J}{s^{2}}-s^{-1}\hat {\cal L}(h_{s}l_{s}).
\end{eqnarray}

The system of equations (14) and (15) comletely defines the behavior of
$N(\tau),h_{s}(\tau)$, and $l_{s}(\tau)$ which, in turn, define the
evolution of spatial particles distribution via Eqs.(13). In the limit
of our interest here $\tau \ll 1$ the first of Eqs.(14) is reduced to

\begin{eqnarray}
\hat {N}=\frac {\hat {h}_{s}}{\sqrt {s}}(1-2e^{-2\sqrt {s}}+\cdots),
\end{eqnarray}

where in expansion of $\tanh\sqrt {s}$ in power-series of $e^{-2\sqrt{s}}$
the leading term $O(e^{-1/\tau})$, which takes into account the
contribution of reflection from the boundary $x=0$, is retained. The second
characteristic time $\tau_{L}=p$, at which the diffusive length $L$ of
particles becomes equal to the system's size, separates two temporal
regions, $\tau \ll p$ and $\tau \gg p$, within which the growth dynamics
of the particles number is qualitatively different. Within $\tau \ll p$ the
$L$ particles duffusion proceeds, actually, into a semi-infinite medium, and
from (14) we have

\begin{eqnarray}
\hat {N} = \frac {\hat {l}_{s}}{\sqrt {sp}}(1-2e^{-2\sqrt {sp}}+\cdots).
\end{eqnarray}

In the opposite limit, $\tau \gg p$, the $L$ particles distribution becomes
practically uniform, and from (14) it follows

\begin{eqnarray}
\hat {N} = \hat {l}_{s}\left(1 - \frac {1}{3}sp + \cdots\right),
\end{eqnarray}

where in expansion of $\tanh\sqrt{sp}$ in power-series of $\sqrt{sp}$
the leading term $O(p/\tau)$ is retained, which in the case of the power
growth $l_{s}\propto \tau^{n}$  leads to the law

\begin{eqnarray}
N= l_{s}\left[1- \frac {n}{3}\left(\frac {p}{\tau}\right) + \cdots\right].
\end{eqnarray}

In what follows, we shall consider the dynamics of crossover from the
adsorption-controlled regime to the diffusion-controlled one first in the
limit of $\tau\ll p$, when $H$ and $L$ particles diffuse into a
semi-infinite medium, then in the limit $p \ll \tau \ll 1$, when $H$
particles diffuse into a semi-infinite medium at a uniform $L$
particles distribution.

\subsection{Diffusion of H and L particles into a semi-infinite medium
($\tau \ll p$).}

On neglecting the reflection from the boundary $x=0$, we have from
(16) and (17)

\begin{eqnarray}
\hat {h}_{s} = \frac {\hat {l}_{s}}{\sqrt {p}}= \hat {N}\sqrt {s}.
\end{eqnarray}

The system of equations (15) and (20) is reduced to the nonlinear integral
equation that cannot be solved in the general form. Our aim will
be, starting with (20), to obtain asymptotics for the solution of (15)-(17)
in the limits of $J^{D}\mid_{s}\simeq J \gg J^{d}$ (adsorption-controlled
regime) and $J^{d}\simeq J\gg J^{D}\mid_{s}$ (diffusion-controlled regime).

\subsubsection{Adsorption-controlled regime (${\tau \ll p,\tau_{q}}$)}

By assuming $N=J\tau$, i.e. by neglecting the contribution of desorption,
from (20) we obtain

\begin{eqnarray}
\nonumber
h_{s}=l_{s}/\sqrt {p}=(2/\sqrt {\pi})J\sqrt {\tau},
\end{eqnarray}

whence it follows that the adsorption-controlled asymptotics has the form

\begin{eqnarray}
\nonumber
h_{s}=\frac{2}{\sqrt {\pi}}J\sqrt {\tau}(1 + \varrho_{H}),
\end{eqnarray}
\begin{eqnarray}
l_{s}=\frac {2}{\sqrt {\pi}}J\sqrt {\tau p}(1 +  \varrho_{L}),
\end{eqnarray}
\begin{eqnarray}
\nonumber
N=J\tau(1 + \varrho_{N}),
\end{eqnarray}

where $\varrho_{i}\ll 1$. Thus, the condition of smallness of desorption
flux is $J^{d}=h_{s}l_{s}=(4/\pi)J^{2}\sqrt {p}\tau\ll J$, whence it
follows that it takes place at $\tau \ll \tau_{q}=1/J\sqrt {p}$.
Substituting the expression for $J^{d}$ into (15), we obtain
$N=J\tau(1-\frac{2}{\pi}(\tau/\tau_{q})+ \cdots)$. Substituting further
this expression into (16) and (17), with an accuracy to the leading terms,
we finally obtain

\begin{eqnarray}
\nonumber
\varrho_{H}= - \frac {8}{3\pi}\left(\frac {\tau}{\tau_{q}}\right)+
2\sqrt {\pi} ierfc \frac {1}{\sqrt {\tau}} + \cdots,
\end{eqnarray}
\begin{eqnarray}
\varrho_{L}= - \frac {8}{3\pi}\left(\frac {\tau}{\tau_{q}}\right)+
2 \sqrt {\pi} ierfc \sqrt {\frac {p}{\tau}}  + \cdots,
\end{eqnarray}
\begin{eqnarray}
\nonumber
\varrho_{N}= - \frac {2}{\pi}\left(\frac {\tau}{\tau_{q}}\right) + \cdots,
\end{eqnarray}

where the function $ierfc(\eta)=e^{-\eta^{2}}/\sqrt {\pi}-\eta erf(\eta)$ at
$\eta \gg 1$ has the asymptotic form
$e^{-\eta^{2}}(1-3/2\eta^{2}+\cdots)/2\sqrt {\pi} \eta^{2}$.

\subsubsection{Diffusion-controlled regime ($\tau_{q}\ll \tau \ll p$).}

In this limit, by neglecting the contribution of the transient region,
i.e. by assuming $\tau_{q}\rightarrow 0$, from (20) and (15) we have

\begin{eqnarray}
\nonumber
h_{s}=l_{s}/\sqrt {p},\quad h_{s}l_{s}=J,
\end{eqnarray}

whence we find $h_{s}=h_{e}p^{-1/4},l_{s}=l_{e}p^{1/4}$ and,
subsequently, according to (20), we obtain
$N=2\sqrt {J/\pi}p^{-1/4}\sqrt {\tau}$ and, hence,
$J^{D}\mid_{s}=J(\frac {\tau_{q}}{\tau})^{1/2}/\sqrt {\pi}$.
Thus, the diffusion-controlled asymptotics has the form

\begin{eqnarray}
\nonumber
h_{s}=h_{e}p^{-1/4}(1+\lambda_{H}),
\end{eqnarray}
\begin{eqnarray}
l_{s}=l_{e}p^{1/4}(1+\lambda_{L}),
\end{eqnarray}
\begin{eqnarray}
\nonumber
N=\frac {2}{\sqrt {\pi}}N_{e}p^{-1/4}\sqrt
{\tau}(1+\lambda_{N}),
\end{eqnarray}

where $\lambda_{i}\ll 1$. Substituting (23) into Eqs. (15), (16), and (17),
we find that, asymptotically, $\lambda_{H}$ and $\lambda_{L}$ are connected
by relationships

\begin{eqnarray}
\nonumber
\lambda_{H}-\lambda_{L}=-2\left(erfc \sqrt {\frac {p}{\tau}} -
erfc \frac {1}{\sqrt {\tau}}\right)+ \cdots,
\end{eqnarray}
\begin{eqnarray}
\nonumber
\lambda_{H}+\lambda_{L}= -\frac {1}{\sqrt {\pi}}\left(\frac {\tau_{q}}
{\tau}\right)^{1/2}+ \cdots,
\end{eqnarray}

whence, with taking account of (16) and (17), we finally obtain

\begin{eqnarray}
\nonumber
\lambda_{H}= -\frac {1}{2\sqrt {\pi}}\left(\frac
{\tau_{q}}{\tau}\right)^{1/2} - erfc \sqrt {\frac {p}{\tau}} +
erfc \frac {1}{\sqrt {\tau}},
\end{eqnarray}
\begin{eqnarray}
\lambda_{L}= -\frac {1}{2\sqrt \pi}\left(\frac {\tau_{q}}{\tau}\right)^{1/2}
+ erfc \sqrt {\frac {p}{\tau}} - erfc \frac {1}{\sqrt {\tau}},
\end{eqnarray}
\begin{eqnarray}
\nonumber
\lambda_{N}= -\frac {\sqrt {\pi}}{4}\left(\frac {\tau_{q}}{\tau}\right)^{1/2}
- \sqrt {\pi}\left(ierfc \sqrt {\frac {p}{\tau}} + ierfc \frac {1}
{\sqrt {\tau}}\right),
\end{eqnarray}

where comlementar error function $erfc(\eta)=1-erf(\eta)$ at $\eta \ll 1$
has the asymptotics $e^{-\eta^{2}}(1-1/2\eta^{2}+\cdots)/\sqrt {\pi}\eta$.

\subsection{Diffusion of H particles into a semi-infinite medium at uniform
distribution of L particles ($ p \ll \tau \ll 1$).}

By neglecting the $H$ particles reflection from the boundary $x=0$ and
nonuniformity of the distribution of $L$ particles, from (16) and (18)
we have

\begin{eqnarray}
\hat {N}=\hat {l}_{s}=\hat {h}_{s}/\sqrt {s}.
\end{eqnarray}

Evidently, the character of the crossover onto regime (25) depends on the
$\tau_{q}/\tau_{L}$ relation. In the region of $\tau_{q} \ll \tau_{L}$
regime (25) is realized after a quasiequilibrium $h_{s}l_{s}\simeq J$ has
been reached, whereas in the opposite limit, $\tau_{L} \ll \tau_{q}$,
the crossover to regime (25) occurs directly at the initial stage when
the desorption can yet be neglected.

\subsubsection{Adsorption-controlled regime ($p \ll \tau \ll \tau_{*}$).}

By assuming $N=J\tau$, i.e. by neglecting the contribution of desorption,
from (25) we have

\begin{eqnarray}
N=l_{s}=J\tau, \quad  h_{s}=(2/\sqrt {\pi})J\sqrt {\tau},
\end{eqnarray}

whence it follows that the adsorption-controlled asymptotics has the form

\begin{eqnarray}
\nonumber
h_{s}=\frac{2}{\sqrt {\pi}}J\sqrt {\tau}(1 + \Delta_{H}),
\end{eqnarray}
\begin{eqnarray}
l_{s}=J\tau(1 + \Delta_{L}),
\end{eqnarray}
\begin{eqnarray}
\nonumber
N=J\tau(1 + \Delta_{N}).
\end{eqnarray}

According to (26), the condition of smallness of the desorption flux takes
now the form $J^{d}=h_{s}l_{s}\simeq (2/\sqrt {\pi})J^{2}\tau^{3/2} \ll J$,
whence it follows that it is realized at times $\tau \ll \tau_{*}=J^{-2/3}$.
So, under the conditions of uniform distribution of $L$ particles, there
appears a new characteristic time scale $\tau_{*}= J^{-2/3}$ in the vicinity
of which the crossover from the adsorption-controlled to the
diffusion-controlled regime occurs. Using the exact series expansion in
particular points

\begin{eqnarray}
\nonumber
\coth\sqrt {sp} =\frac {1}{\sqrt {sp}} + 2\sum_{n=1}^{\infty}\frac
{\sqrt{sp}}{sp +(\pi n)^{2}},
\end{eqnarray}

we find from (14)

\begin{eqnarray}
\hat {l}_{s}=\hat {N}\left\{1 + \frac {1}{3} sp - 2\sum_{n=1}^{\infty}
\frac {(sp/\pi n)^{2}}{sp + (\pi n)^{2}}\right\},
\end{eqnarray}

and, therefore, in the limit $\tau/\tau_{*}\rightarrow 0$ we have exactly

\begin{eqnarray}
\nonumber
l_{s}=J\tau\left\{1 +\frac {1}{3}\left(\frac {p}{\tau}\right)\left[1-
\frac {6}{\pi^{2}}\sum_{n=1}^{\infty} n^{-2}e^{-\frac {(\pi n)^{2}\tau}
{p}}\right]\right\},
\end{eqnarray}

whence it is seen that the addition to (19) becomes negligibly small
$O(e^{-\pi^{2}})$ already at $\tau\sim p$. Substituting the expression for
$J^{d}$ into (15), we obtain
$N=J\tau (1- \frac {4}{5\sqrt {\pi}}(\frac {\tau}{\tau_{*}})^{3/2}+\cdots)$.
Substituting then this expression into (16) and (28), respectively, we
finally find

\begin{eqnarray}
\nonumber
\Delta_{H}= - \frac {3\sqrt {\pi}}{8}
\left(\frac {\tau}{\tau_{*}}\right)^{3/2}+
2\sqrt {\pi}ierfc\frac {1}{\sqrt {\tau}}+\cdots,
\end{eqnarray}
\begin{eqnarray}
\Delta_{L}= \frac {1}{3}\left(\frac {p}{\tau}\right)- \frac {4}{5\sqrt {\pi}}
\left(\frac {\tau}{\tau_{*}}\right)^{3/2} + \cdots,
\end{eqnarray}
\begin{eqnarray}
\nonumber
\Delta_{N}= - \frac {4}{5\sqrt {\pi}}
\left(\frac {\tau}{\tau_{*}}\right)^{3/2} + \cdots.
\end{eqnarray}

\subsubsection{Diffusion-controlled regime ($p,\tau_{*} \ll \tau \ll 1$)}

By neglecting the contribution of the transient region, i.e. by taking
$p,\tau_{*}\rightarrow 0$ and assuming that $h_{s}$ changes by the
power law $h_{s}\sim \tau^{n}$, from (25) and (15) we have

\begin{eqnarray}
N=l_{s}=\gamma(n)h_{s}\sqrt {\tau}, \quad h_{s}l_{s}=J,
\end{eqnarray}

where $\gamma(n)=\Gamma(n+1)/\Gamma(n+3/2)$ and $\Gamma(n)$ is the gamma
function. From (30) it immediately follows $n=-1/4$. Thus, the exact
asymptotic solution of (15),(25) in the limit
$p\rightarrow 0, J\rightarrow \infty$ is

\begin{eqnarray}
N=l_{s}=l_{e}\beta\tau^{1/4}, \quad h_{s}=h_{e}\beta^{-1}\tau^{-1/4},
\end{eqnarray}

where $\beta =[\Gamma(3/4)/\Gamma(5/4)]^{1/2}=1.1627366...$.
With account taken of the contribution of the transient region and
reflection of $H$ particles from the boundary $x=0$, we, finally, obtain
the diffusion-controlled asymptotics in the form

\begin{eqnarray}
\nonumber
h_{s}=h_{e}\beta^{-1}\tau^{-1/4}(1+\phi_{H}),
\end{eqnarray}
\begin{eqnarray}
l_{s}=l_{e}\beta\tau^{1/4}(1+\phi_{L}),
\end{eqnarray}
\begin{eqnarray}
\nonumber
N=N_{e}\beta\tau^{1/4}(1+\phi_{N}).
\end{eqnarray}

Substituting (32) into (15) and (18) and taking $\phi_{i}\ll 1$ we find
that, asymptotically, $\phi_{H},\phi_{L}$, and $\phi_{N}$ are connected by
the relationships

\begin{eqnarray}
\phi_{H}+\phi_{L}= - \frac {\beta}{4}
\left(\frac {\tau_{*}}{\tau}\right)^{3/4}+\cdots
\end{eqnarray}

and

\begin{eqnarray}
\phi_{N}=\phi_{L} - \frac {1}{12}\left(\frac {p}{\tau}\right)+\cdots.
\end{eqnarray}

Substituting then (32) into Eq.(16), we come to the equation

\begin{eqnarray}
\beta^{2}\hat {\cal L}(\phi_{N}\tau^{1/4})=s^{-1/2}\hat {\cal L}
(\phi_{H}/\tau^{1/4}) - \frac {2\Gamma(3/4)}{s^{5/4}}e^{-2\sqrt {s}},
\end{eqnarray}

which, on neglecting the contribution of $H$ particles reflection from
the boundary $x=0$, takes the form

\begin{eqnarray}
\beta^{2}\hat {\cal L}(\tilde {\phi}_{N}\tau^{1/4})=s^{-1/2}\hat {\cal L}
(\tilde {\phi}_{H}/\tau^{1/4}).
\end{eqnarray}

Assuming that $\tilde {\phi}_{i}$ decay asymptotically in time by the
power-law

\begin{eqnarray}
\tilde {\phi}_{i} = {\cal C}_{i}/\tau^{\nu_{i}} + \cdots,
\end{eqnarray}

where ${\cal C}_{i}(p,\tau_{*})\rightarrow 0$ as $p,\tau_{*}\rightarrow 0$,
from (33),(34) and (36) it can easily be shown that the exponents
$\nu_{i}\geq 3/4$. Indeed, let $\nu_{H} < 3/4$. Then, from (36) it follows
that $\nu_{L}=\nu_{H}$, and the coefficients ${\cal C}_{L}$ and
${\cal C}_{H}$ have like signs, which contradicts the condition (33).
Thus, $\nu_{i} \geq 3/4$ which suggests that in contrast to the
diffusion-controlled asymptotics (23), the sign and value of the
${\cal C}_{L}$ and ${\cal C}_{H}$ are governed by all the prehistory of
transition to the asymptotics (31) and can be calculated only numerically.
In the next sections we shall discuss the behavior of $\tilde {\phi}_{L}$
and $\tilde {\phi}_{H}$ in more detail and demonstrate that it is quite
nontrivial. We now consider the contribution into $\phi_{i}$ due to the
reflection of $H$ particles from the boundary $x=0$. According to (33) and
(34), in the limit $p,\tau_{*} \rightarrow 0$ we have

\begin{eqnarray}
\nonumber
\phi_{H}= -\phi_{L}=-\phi_{N}=\phi^{(0)}.
\end{eqnarray}

Substituting $\phi^{(0)}$ into Eq.(35), we obtain

\begin{eqnarray}
\beta^{2}\hat {\cal L}(\phi^{(0)}\tau^{1/4})=
\frac {2\Gamma(3/4)}{s^{5/4}}e^{-2\sqrt {s}} -
s^{-1/2}\hat {\cal L}(\phi^{(0)}/\tau^{1/4}).
\end{eqnarray}

It is easy to check that the inverse Laplace transform of the solution of
Eq.(38) ought to have the form

\begin{eqnarray}
\phi^{(0)}=\frac {2\Gamma(5/4)}{\sqrt {\pi}}\tau^{3/4}e^{-1/\tau}(1+g),
\end{eqnarray}

where

\begin{eqnarray}
\nonumber
g=c_{1/2}\tau^{1/2}+c_{1}\tau +c_{3/2}\tau^{3/2} + \cdots.
\end{eqnarray}

Substituting Eq.(39) into (38) and equlizing the coefficients at the
same $s$ powers, we finally find

\begin{eqnarray}
\nonumber
c_{1/2}=-\frac {1}{\beta^{2}}, c_{1}=-\frac {15}{16}+\frac {1}{\beta^{4}},
c_{3/2}=\frac {1-c_{1}}{\beta^{2}}.
\end{eqnarray}

\subsection{Two characteristic paths of the crossover to the
diffusion-controlled regime with uniform L particle distribution.}

According to subsections A) and B), depending on the relation of three
characteristic times $\tau_{L}=p,\tau_{q}$, and $\tau_{*}$, the crossover to
the diffusion-controlled regime with uniform $L$ particles distribution
proceeds by one of the two qualitatively different scenarios. One of these is
realized in the limit $\tau_{q}\ll p \ll 1$, when at first, a quasiequilibrium
between the input and desorption fluxes is established in the system
(crossover $A1\rightarrow A2$ at $\tau\sim \tau_{q}$), following which
the $L$ particles distribution becoms uniform (crossover $A2\rightarrow B2$
at $\tau \sim p$). In this case, the surface concentrations $h_{s},l_{s}$ and
particle number $N$ go, respectively, through the following chains of
power-law asymptotics:

\begin{eqnarray}
\nonumber
h_{s}:\quad \tau^{1/2}\rightarrow\tau^{0}\rightarrow \tau^{-1/4},
\end{eqnarray}
\begin{eqnarray}
l_{s}:\quad \tau^{1/2}\rightarrow\tau^{0}\rightarrow\tau^{1/4},
\end{eqnarray}
\begin{eqnarray}
\nonumber
N:\quad \tau^{1}\rightarrow \tau^{1/2}\rightarrow \tau^{1/4}.
\end{eqnarray}

The second scenario is realized in the limit $p\ll\tau^{*}\ll 1$, when at
first the $L$ particles distribution becomes uniform (crossover
$A1 \rightarrow B1$ at $\tau\sim p$), whereupon a quasiequilibrium is
established between the input and desorption fluxes (crossover
$B1\rightarrow B2$ at $\tau\sim\tau_{*}$). In this case, $h_{s},l_{s}$, and
$N$ go, respectively, through the following chains of power-law asymptotics:

\begin{eqnarray}
\nonumber
h_{s}:\quad \tau^{1/2}\rightarrow \tau^{-1/4},
\end{eqnarray}
\begin{eqnarray}
l_{s}:\quad \tau^{1/2}\rightarrow \tau^{1}\rightarrow \tau^{1/4},
\end{eqnarray}
\begin{eqnarray}
\nonumber
N:\quad \tau^{1}\rightarrow \tau^{1/4}.
\end{eqnarray}

It can easily be shown from Eqs.(13) that at each of the given power-law
asymptotics, $h_{s}\propto \tau^{n},\tau \ll 1$, the spatial distribution
of $H$-particles changes, on neglecting the contribution of the transient
regions, by the law

\begin{eqnarray}
h(x,\tau)=h_{s}{\cal F}_{n}\left(\frac {1-x}{2\sqrt {\tau}}\right),
\end{eqnarray}

where

\begin{eqnarray}
{\cal F}_{n}(\xi)=
\left\{\begin{array}{ll}
\pi^{1/2} ierfc(\xi), \quad &n=1/2 \nonumber\\
\\
erfc(\xi), \quad &n=0 \nonumber\\
\\
\frac {2^{1/4}\Gamma(3/4)}{\sqrt \pi}e^{-\xi^{2}/2}{\cal D}_{-1/2}
(\sqrt {2}\xi), \quad &n=-1/4.\nonumber\\
\end{array}
\right.
\end{eqnarray}

Here ${\cal D}_{\nu}(z)$ is the function of parabolic cylinder which
at $z \gg 1$ has the asymptotics ${\cal D}_{\nu}(z)=e^{-z^{2}/4}z^{\nu}
[1- \nu(\nu-1)/2z^{2} + \ldots]$. The spatial distribution of $L$ particles
at asymptotics $l_{s}\propto \tau^{n},\tau \ll p$ with $n=1/2,0$ has exactly
the same form, except for the substitution $\tau \rightarrow \tau/p$.

By comparing the paired ratios of the characteristic times
$\tau_{L},\tau_{q}$, and $\tau_{*}$, we find

\begin{eqnarray}
\frac {\tau_{L}}{\tau_{q}}=
\left(\frac {\tau_{L}}{\tau_{*}}\right)^{3/2}=
\left(\frac {\tau_{*}}{\tau_{q}}\right)^{3}={\cal K}=p^{3/2}J,
\end{eqnarray}

whence it is seen that the character of the system's evolution is governed
by the value of the parameter ${\cal K}=p^{3/2}J$: at any $p$ and $J$
relation in the limit ${\cal K}\gg 1$ the system evolves in accord with
the chains of asymptotics (40), whereas in the opposite limit,
${\cal K} \ll 1$, the system evolves in accord with the chains of asymptotics
(41). On neglecting the reflection of $H$ particles from the boundary $x=0$,
one can easily see that in the limit
${\cal K}\rightarrow \infty (\tau_{q}/p \rightarrow 0)$ the only
characteristic time scale is defined by the quantity $p$, therefore,
as suggested by (23) and (32), the behavior of $h_{s}$ and $l_{s}$ should
have the scaling form

\begin{eqnarray}
h_{s}/h_{e}=p^{-1/4}w_{H}(\tau/p), \quad l_{s}/l_{e}=p^{1/4}w_{L}(\tau/p),
\end{eqnarray}

where the scaling functions $w_{i}(\zeta)$ have asymptotics
$w_{H,L}(\zeta)=1 \mp erfc(1/\sqrt{\zeta})$ at
$\zeta \ll 1$ and $w_{H,L}(\zeta)\simeq(\beta \zeta^{1/4})^{\mp 1}$ at
$\zeta \gg 1$. Indeed, substituting (44) into (14) and (15) and going to the
Lasplace transform with respect to the reduced time
$\zeta=\tau/p \rightarrow s$, in the limit ${\cal K}\rightarrow \infty$
we find

\begin{eqnarray}
\nonumber
w_{H}w_{L}=1, \quad \hat {w}_{H}=\hat {w}_{L}\tanh \sqrt {s},
\end{eqnarray}

whence it immediately follows $w_{i}=w_{i}(\tau/p)$ and, hence,
$\tilde {\phi}_{i}=\tilde {\phi}_{i}(\tau/p)$. In the opposite limit,
${\cal K} \rightarrow 0 (p/\tau_{*} \rightarrow 0)$, the only characteristic
time scale is defind by the quantity $\tau_{*}$, therefore, in accord with
(27) and (32), the behavior of $h_{s}$ and $l_{s}$ should have the scaling
form

\begin{eqnarray}
h_{s}/h_{e}=J^{1/6}v_{H}(\tau/\tau_{*}), \quad
l_{s}/l_{e}=J^{-1/6}v_{L}(\tau/\tau_{*}),
\end{eqnarray}

where the scaling functions $v_{i}(\zeta)$ have asymptotics
$v_{H}(\zeta)\simeq 2\sqrt {\zeta/\pi}, v_{L}(\zeta)\simeq \zeta$ at
$\zeta \ll 1$ and $v_{H,L}\simeq (\beta \zeta^{1/4})^{\mp 1}$ at
$\zeta \gg 1$. Indeed, substituting (45) into (14) and (15) and going to
the Laplace transform with respect to the reduced time
$\zeta=\tau/\tau_{*} \rightarrow s$, in the limit ${\cal K} \rightarrow 0$
we find

\begin{eqnarray}
\nonumber
\hat {\cal L}(v_{H}v_{L})=s^{-1}(1-s^{2}\hat {v}_{L}), \quad
\hat {v}_{H}=\hat {v}_{L}\sqrt {s},
\end{eqnarray}

whence it immediately follows $v_{i}=v_{i}(\tau/\tau_{*})$ and, hence,
$\tilde {\phi}_{i}=\tilde {\phi}_{i}(\tau/\tau_{*})$.

From (45) it follows that in the limit ${\cal K}\rightarrow 0$ the crossover
to the diffusion-controlled regime is characterized by a comparatively sharp
maximum of $h_{s}/h_{e}$, the height of which depends solely on $J$ and
changes with growing $J$ by the law

\begin{eqnarray}
h_{s}^{M}/h_{e}=b_{H}J^{1/6}.
\end{eqnarray}

In the opposite limit ${\cal K}\rightarrow \infty$, according to (23),(44),
a maximum of $h_{s}/h_{e}$ is degenerated to an extended plateau, of which
the height depends on $p$ alone

\begin{eqnarray}
h_{s}^{M}/h_{e}=p^{-1/4},
\end{eqnarray}

and the extension, defined as the ratio of the times, which bound the region
$\lambda_{H}\leq \epsilon \ll 1$, grows with $J$ as
$\propto {\cal K}\epsilon^{2}/\ln {(1/\epsilon)}$.

In the next subsections we give the results of a detailed numerical study
of the dynamics of the system in question, enabling one to gain a complete
picture of evolution of $h_{s},l_{s}$, and $N$ "trajectories" between two
limiting regimes (40) and (41).

\subsection{Numerical calculations}

The numerical integration of equations (6) was performed by means of the
{\it implicit} discretization scheme of increased accuracy with an additional
{\it "fictitious"} node at the surface. The scheme allowed performing the
calculations in the system with strong difference in species diffusivities
with an accuracy down to $10^{-3}\%$ (see below). The space and time steps
were changed within the ranges
$\delta x\simeq 3\times10^{-4}\div3\times10^{-6}$ and
$\delta \tau \simeq 10^{-4}\div10^{-11}$, respectively, with the number
of time steps being $10^{5}\div10^{6}$. The evolution dynamics of
$h_{s}(\tau),l_{s}(\tau)$, and $N(\tau)$ was studied in the ranges of
$p=10^{-10}\div1$ and $J=10^{-1}\div10^{12}$.

\subsubsection{Kinetic diagrams $p-J-\tau$}

Fig.5(a) illustrates the time dependences $h_{s}(\tau)/h_{e},
l_{s}(\tau)/l_{e}$, and $N(\tau)/N_{e}$, calculated numerically at a fixed
flux density $J=10^{8}$ for $p$ values ranging from $1$ to $10^{-9}$
(the arrows show the directions of shift of the corresponding trajectories
at a change of $p$ from $p=1$ (bold curves) to $p=10^{-9}$ (bold curves)).
One can clearly see the whole set of power-law asymptotic regions (40),(41)
of $h_{s},l_{s}$, and $N$ trajectories and the character of their evolution
as $p$ (and, hence, ${\cal K}$) are decreased from the values corresponding
to the limiting regime (40) ($p \gg 5\times10^{-6}, {\cal K} \gg 1$) to those
of (41) ($p \ll 5\times 10^{-6}, {\cal K} \ll 1$) where the dependences of
the trajectories on $p$ dissapears and they asymptotically approach those
shown in bold lines. In accord with (46) and (47), as $p$ decreases, the
plateau on the curves $h_{s}(\tau)/h_{e}$ is transformed to a comparatively
sharp maximum, the height of which depends only on $J$. At all $p \ll 1$ the
$h_{s}/h_{e},l_{s}/l_{e}$, and $N/N_{e}$ trajectories come to the universal
asymptotics $(\beta\tau^{1/4})^{\mp 1}$ (31). Herewith, as seen from Fig.5a
and will be demonstrated in detail in what follows, the $h_{s}(\tau)/h_{e}$
trajectories always come to the asymptotics $(\beta\tau^{1/4})^{-1}$ only
{\it from above}, whereas the $l_{s}(\tau)/l_{e}$ trajectories always come to
the asymptotics $\beta\tau^{1/4}$ only {\it from below}.

By defining the boundaries of the asymptotic regions so that within their
confines the condition

\begin{eqnarray}
\max\{|\delta_{a}|,|d\delta_{a}/d\ln \tau|\} \leq 0.01
\end{eqnarray}

shall hold (here $\delta_{a}\equiv \{\varrho_{i}, \lambda_{i}, \Delta_{i},
\phi_{i}, \sigma_{i}\}, \sigma_{i}$ describes the exponential relaxation to
the steady state according to Eqs.(69)), we have carried out an extensive
numerical study of the positions of the corresponding boundaries for
$J=10^{8}$ and $p$ values, ranging from $1$ to $10^{-10}$ (we have
calculated and analyzed trajectories for 100 values of $p$, 10 for each
order). Figs.5b and c show the kinetic $p - \tau$ diagrams of the regions
of the power-law asymptotics and steady state (shaded by light gray) for the
$h_{s}(\tau)/h_{e}$ and $l_{s}(\tau)/l_{e}$ trajectories, respectively.
In Fig.5c by dark gray is distinguished the $\beta\tau^{1/4}$ asymptotic
region for the $N(\tau)/N_{e}$ trajectories of the growth of the particles
number. From Figs.5b,c it is seen that the dashed lines of the characteristic
times $\tau_{q},\tau_{*},\tau_{L}=p$ and $\tau_{H}=1$ divide the $p-\tau$
plane into several segments, confining the regions of the power-law
asymptotics $\tau^{n}$ with $n=1/2,0,1$ and $\mp1/4$. The boundaries of these
regions go in parallel with the lines of the corresponding characteristic
times, in accord with Eqs. (22), (24), (29), (39), (44) and (45). The
$\tau_{q},\tau_{L}$, and $\tau_{*}$ lines intersect in the point ${\cal K}=1$
($\tau_{q}=\tau_{L}=\tau_{*}=4.64158\times 10^{-6}$), shown by filled circle,
in accord with Eq.(43). As the $n=1/2,0,1$ region boundary positions are
described by Eqs. (22), (24) and (26), we shall focus mainly on
the $n=\mp1/4$ region boundary positions.

From Figs.5b,c it is seen that, in accord with (44), at
$p \gg 5\times10^{-6}({\cal K} \gg 1)$ the left-hand boundaries of the
$n=\mp1/4$ regions go in parallel with the $\tau_{L}$ line down to the point
of intersection of left- and right-hand boundaries of $n=0$ regions
($p\approx 2\times10^{-3}, {\cal K}\approx 5\times10^{3}$) where the plateau
on the $h_{s}$ and $l_{s}$ trajectories disappears. Above this point the
$h_{s}/h_{e}$ and $l_{s}/l_{e}$ approarch the $(\beta\tau^{1/4})^{\mp1}$
asymptotics {\it simultaneously} and, which is important, this occurs
at $\tau\approx p$, i.e. {\it long before} the $L$ particles distribution in
the bulk becomes uniform (left boundary of the $n=1/4$ region shaded by dark
gray in Fig.5c). Such synhchronization of the trajectories directly
follows from Eqs.(33) and (44), according to which in the scaling limit
${\cal K}\rightarrow \infty$ ($\tau_{*}/p\rightarrow 0$) $w_{H}w_{L}=1$ and
$\phi_{H}=-\phi_{L}$. In the opposite limit $p\ll 5\times10^{-6}
({\cal K} \ll 1)$ the left boundaries of $n=\mp1/4$ regions become parallel
to the $\tau_{*}$ line in the vicinity of the intersection point of $n=1/2$
boundary (Fig.5b) with the $\tau_{L}$ line ($p\approx 3\times10^{-7},
{\cal K}\approx2\times10^{-2}$), in accord with Eq.(45). Below this point
(i) the behavior of $h_{s}(\tau)$ obeys the scaling law (45), i.e. it ceases
to be dependent on $p$ and (ii) the boundaries of $n=1/4$ regions of growth
of $l_{s}(\tau)$ and $N(\tau)$ merge, i.e. by the instant of the transition
to asymptotics $\beta\tau^{1/4}$ the distribution of $L$ particles becomes
{\it uniform}. Essentially, that in this region of $p$ and ${\cal K}$ the
behavior of $h_{s}$ and $l_{s}$ becomes strongly {\it asymmetric}: the
transition of the $h_{s}/h_{e}$ trajectories to $(\beta\tau^{1/4})^{-1}$
asymptotics takes place {\it well before} the $l_{s}/l_{e}$ trajectories
reach the $\beta\tau^{1/4}$ asymptotics. Below the point
$p\approx 2\times10^{-9}, {\cal K}\approx10^{-5}$, where on the kinetic
diagram of Fig.5c there appears the $n=1$ region, the merging of the
$l_{s}(\tau)$ and $N(\tau)$ trajectories takes place already on the stage of
the adsorption-controlled regime, beyond which the $l_{s}(\tau)$ trajectory
ceases to be $p$-independent.

With account taken of the fact that the boundaries of the asymptotic regions
are strongly related to the characteristic times $\tau_{L}, \tau_{q}$, and
$\tau_{*}$, one can easily get from Fig.5b,c the idea of how the kinetic
diagrams evolve with changing density of the input flux $J$. Indeed, as $J$
grows, the point ${\cal K}=1$ should shift along the $\tau_{L}$ line, as
shown by arrows in Fig.5b,c, entailing the $\tau_{q}$ and $\tau_{*}$ lines.
The $\tau_{q}$ and $\tau_{*}$ lines, shifting in parallel to themselves,
should in turn entail the lines of the corresponding boundaries of the
asymptotic regions, so that with growing $J$ the $n=0,\mp1/4$ regions should
expand in a self-similar manner, and the $n=1/2,1$ regions contract in a
self-similar manner. From the conditions $\tau_{-1/4}\simeq\tau_{r}$ and
$\tau_{1/4}\simeq\tau_{r}$, where $\tau_{-1/4}=5.855\tau_{*}$ and
$\tau_{1/4}=93.88\tau_{*}$ are the positions of the left boundaries of
regions $n=-1/4$ and $n=1/4$ at ${\cal K}\rightarrow 0$
and $\tau_{r}\approx 0.2$ is the position of their right boundary associated
with the influence of the $H$ particles reflection from the boundary $x=0$,
one can conclude that the starting flux densities for the appearance of
$\tau^{-1/4}$ and $\tau^{1/4}$ asymptotics are $J_{-1/4}\approx 10^{2}$
and $J_{1/4}\approx 10^{4}$, respectively. From Fig.5b,c it follows that
the vertices of $n=\mp1/4$ regions for the surface concentrations and the
number of particles are in the points $p_{s}\approx 0.1$ and
$p_{N}\approx 2\times10^{-2}$, respectively. Thus, the $J-p$ limits of
$(\beta\tau^{1/4})^{\mp1}$ asymptotics are defined by the following
conditions:

\begin{eqnarray}
\nonumber
h_{s}: p < p_{s}\approx 0.1, J > J_{-1/4}\approx 10^{2},
\end{eqnarray}
\begin{eqnarray}
\nonumber
l_{s}: p < p_{s}\approx 0.1, J > J_{1/4}\approx 10^{4},
\end{eqnarray}
\begin{eqnarray}
\nonumber
N: p < p_{N}\approx 2\times10^{-2}, J > J_{1/4}\approx 10^{4}.
\end{eqnarray}

\subsubsection{Maximum of $h_{s}$}

Figs.6a and b illustrates the behavior of the maximum of the surface $H$
particles concentration, $h_{s}^{M}/h_{e}$,(a) and the time for which
this maximum is reached, $\tau_{M}$,(b) as a function of growing $J$.
The curves are calculated numerically at fixed $p=10^{-1}, 10^{-2}, 10^{-3},
10^{-4}, 10^{-5}$ and $10^{-6}$. It is seen that, in accord with (45),(46),
at small ${\cal K} \ll 1$ the calculated curves approach the $p$-independent
power-law asymptotics (shown in the dashed lines)

\begin{eqnarray}
h_{s}^{M}/h_{e}=0.7221 J^{1/6}, \quad \tau_{M}= 1.125 J^{-2/3}.
\end{eqnarray}

As $J$ grows, the dependences of $h_{s}^{M}/h_{e}$ and $\tau_{M}$ on $J$
deviate from the $p$-independent asymptotics (49) the earlier the large is
$p$, reaching at ${\cal K} \gg 1$ the $J$-independent plateau (47) (as seen
for the curves for $p=10^{-1}$ and $p=10^{-2}$). Interestingly that at
$p \ll 1$ the curves approach the power-law asymptotics (49) already at
comparatively small flux densities $J\approx 50$, exceeding $J_{c}$ by
no more than an order of magnitude.

\subsubsection{Behavior of $\phi_{H}$ and $\phi_{L}$}

We shall now consider the regularities of the behavior of transient terms
$\phi_{H}(\tau)$ and $\phi_{L}(\tau)$, which characrerize the kinetics of the
transition of surface concentrations to the universal asymptotics
$(\beta\tau^{1/4})^{\mp1}$. Fig.7 shows the dependences $\phi_{H}(\tau)$(a)
and $\phi_{L}(\tau)$(b), calculated numerically at $p=10^{-4}$ for $J$
values, ranging from $J=10^{4}$ to $J=10^{12}$ (from ${\cal K}=10^{-2}$ to
${\cal K}=10^{6}$, respectively). It is seen that (i) the $\phi_{H}$ value
at any ${\cal K}$ first crosses zero, changing the sign from - to + , then
reaches a maximum and asymptotically approaches zero only {\it from above};
(ii) the $\phi_{L}$ value at any ${\cal K}$ asymptotically approaches zero
only {\it from below}, in this case, with ${\cal K}< 1$ it occurs
monotonously whereas with ${\cal K} \gg 1$ the $\phi_{L}$ value first crosses
zero, changing the sign from + to - , then reaches in modulus a maximum, and
only after this begins to approach zero; (iii) with a growth in
$J$ the $\phi_{H}$ maximum is shifted left and its amplitude grows, whereas
the $|\phi_{L}|$ maximum is shifted right and its amplitude drops, so that at
${\cal K} > 10^{6}$ the behavior of $\phi_{H}(\tau)$ and $\phi_{L}(\tau)$
becomes comletely "symmetric", in agreeement with (33), (44)

$$
\phi_{H}(\tau)=|\phi_{L}(\tau)|.
$$

From Fig.8 where are presented the sections of the dependences
$\phi_{H}(\tau)> 0$(a) and $\phi_{L}(\tau)< 0$(b) replotted in double
logarithmic coordinates, we find that at $J \gg J_{1/4}$ the
$\phi_{H}$ and $\phi_{L}$ values decrease at a sufficient distance from
the $\phi_{H}$ maximum by the power law (37) with the exponents
$\nu_{H}=\nu_{L}=3/4$

\begin{eqnarray}
\tilde{\phi}_{i}={\cal C}_{i}/\tau^{3/4} + \cdots
\end{eqnarray}

up to $\tau\approx 0.1\div 0.2$ where a rapid growth of $|\phi_{i}|$ begins
due to the reflection of $H$ particles from the boundary $x=0$.
At the growth stage the $|\phi_{i}|$ dependences are seen to go exactly onto
the bold $\phi^{(0)}$ curve, calculated from Eq.(39).
It is important to note that from Fig.8 it directly follows that the error
of numerical calculations does not exceed $\approx10^{-3}\%$.
According to Eq.(44) in the scaling limit ${\cal K}\rightarrow \infty$
from (50) and (33) it follows

$$
\tilde{\phi}_{i} = m^{\infty}_{i}(p/\tau)^{3/4}+\cdots,
$$

where the coefficients $m^{\infty}_{H}=-m^{\infty}_{L}$ are independent
of $J$ and $p$. From the data of Fig.8 we find at ${\cal K}>10^{6}$

\begin{eqnarray}
m^{\infty}_{H}= -m^{\infty}_{L}= 0.023
\end{eqnarray}

and obtain for the maximum $|\phi_{i}|^{m}$ and the time of its attaining
$\tau^{m}_{i}$

\begin{eqnarray}
\phi^{m}_{H}=|\phi_{L}|^{m}=0.00798,\quad \tau^{m}_{i}=2.51p.
\end{eqnarray}

In Fig.9 are plotted in double logarithmic coordinates the sections of
the dependences $\phi_{H}(\tau)>0$(a) and $\phi_{L}(\tau)<0$(b), calculated
numerically at a fixed flux density $J=10^{6}$ for the $p$ values, ranging
from $p=10^{-2}$ to $p=10^{-6}$ (from ${\cal K}=10^{3}$ to
${\cal K}=10^{-3}$, respectively). The both maxima, $\phi_{H}^{M}$ and
$|\phi_{L}|^{M}$, are seen to shift towards smaller $\tau$ with decreasing
$p$, herewith the $\phi_{H}$ maximum amplitude decreases, whereas the
$|\phi_{L}|$ maximum amplitude increases, and at ${\cal K}<1$ it disappears.
At $p \ll p_{s}$ the $\phi_{H}(\tau)$ and $\phi_{L}(\tau)$ dependences
decay at a distance from $\phi_{H}^{m}$ by the power law (50), converging to
the $p$ independent trajectories at
$p\approx 10^{-6}({\cal K}\approx 10^{-3})$.
According to Eq.(45) in the scaling limit ${\cal K}\rightarrow 0$
from (50) and (33) it follows

$$
\tilde{\phi}_{i}=m^{0}_{i}(\tau_{*}/\tau)^{3/4}+\cdots,
$$

where the $p$- and $J$-independent coefficients $m^{0}_{i}$ should satisfy
the condition

\begin{eqnarray}
m^{0}_{H}+m^{0}_{L}=-\beta/4=-0.29068... .
\end{eqnarray}

From the data of Fig.9, in accord with (53), we find at ${\cal K}< 10^{-3}$

\begin{eqnarray}
m^{0}_{H}=0.014, \quad m^{0}_{L}=-0.305
\end{eqnarray}

and obtain for the $\phi^{m}_{H}$ maximum and the time of its attaining
$\tau^{m}_{H}$

\begin{eqnarray}
\phi^{m}_{H}=0.00108, \quad \tau^{m}_{H}=17.3\tau_{*}.
\end{eqnarray}

\subsection{Scaling and universality}

So far the main attention has been focused on the limiting regimes
${\cal K}\ll 1 $ and ${\cal K}\gg 1$ and on specific features of the
crossover between these regimes at a change in $J$ for $p=$const. or at
a change in $p$ for $J=$const. We shall now show that at $\tau \ll 1$, when
the $H$ particles reflection from the boundary $x=0$ can be neglected, the
behavior of $h_{s},l_{s}$, and $N$ is described by the scaling laws

\begin{eqnarray}
\nonumber
h_{s}=h_{e}p^{-1/4}W_{H}({\cal K},T),
\end{eqnarray}
\begin{eqnarray}
l_{s}=l_{e}p^{1/4}W_{L}({\cal K},T),
\end{eqnarray}
\begin{eqnarray}
\nonumber
N=N_{e}p^{1/4}W_{N}({\cal K},T),
\end{eqnarray}

where $T=\tau/p$ and the scaling functions $W_{i}({\cal K},T)$ depend on
the only parameter ${\cal K}=p^{3/2}J$. Indeed, going to the Laplace
transform with respect to the reduced time $T\rightarrow s$, on
substituting (56) into Eqs.(14) and (15) and neglecting the terms, related
to the $H$ particles reflection from the boundary $x=0$, we easily find

\begin{eqnarray}
\hat {W}_{N}= \frac {\hat {W}_{H}}{\sqrt {s}}=\frac {\hat {W}_{L}}
{\sqrt {s}}\tanh \sqrt {s},
\end{eqnarray}
\begin{eqnarray}
\hat {W}_{N}=\frac {\sqrt {\cal K}}{s^{2}}(1 - s\hat {\cal L}(W_{H}W_{L})),
\end{eqnarray}

whence it follows $W_{i}=W_{i}({\cal K},T)$. In agreement with the results
of Sections IVA and B, the asymptotic behavior of the scaling functions
$W_{i}({\cal K},T)$ on the power-law portions of the $h_{s},l_{s}$, and $N$
trajectories have the following form:

A) Diffusion of $H$ and $L$ particles into a semi-infinite medium
($T \ll 1$).

{\it 1. Adsorption-controlled regime} ($T \ll 1,1/{\cal K}$).

\begin{eqnarray}
\nonumber
W_{H}=\frac {2}{\sqrt {\pi}} \sqrt {{\cal K}T}
\left(1-\frac {8}{3\pi} {\cal K}T + \cdots\right),
\end{eqnarray}
\begin{eqnarray}
W_{L}=\frac {2}{\sqrt {\pi}} \sqrt {{\cal K}T}
\left(1-\frac {8}{3\pi} {\cal K}T +
Te^{-1/T} + \cdots\right),
\end{eqnarray}
\begin{eqnarray}
\nonumber
W_{N}=T\sqrt {\cal K}\left(1 - \frac {2}{\pi}{\cal K}T + \cdots\right).
\end{eqnarray}

{\it 2. Diffusion-controlled regime} ($1/{\cal K}\ll T \ll 1$).

\begin{eqnarray}
W_{H,L}=1 - \frac {1}{2\sqrt {\pi {\cal K}T}} \mp erfc\frac {1}{\sqrt {T}}
+ \cdots,
\end{eqnarray}
\begin{eqnarray}
\nonumber
W_{N}=2\sqrt {T/\pi}\left(1 - \case 1/4 \sqrt {\frac {\pi}{{\cal K}T}} -
\sqrt {\pi} ierfc \frac {1}{\sqrt {T}} + \cdots\right).
\end{eqnarray}

B) Diffusion of $H$ particles into a semi-infinite medium at uniform
distribution of $L$ particles ($1 \ll T \ll 1/p$).

{\it 1. Adsorption-controlled regime} ($1 \ll T \ll 1/{\cal K}^{2/3}$).

\begin{eqnarray}
\nonumber
W_{H}=\frac {2}{\sqrt {\pi}}\sqrt {{\cal K}T}
\left(1 - \frac {3\sqrt {\pi}}{8}{\cal K}T^{3/2}
+ \cdots\right),
\end{eqnarray}
\begin{eqnarray}
W_{L}=T\sqrt {{\cal K}}\left(1 + \frac {1}{3T} - \frac {3\sqrt {\pi}}{8}
{\cal K}T^{3/2} + \cdots\right),
\end{eqnarray}
\begin{eqnarray}
\nonumber
W_{N}=T\sqrt {{\cal K}}
\left(1 - \frac {4}{5\sqrt {\pi}}{\cal K}T^{3/2} + \cdots\right).
\end{eqnarray}

{\it 2. Diffusion-controlled regime} ($T \gg 1, 1/{\cal K}^{2/3}$).

\begin{eqnarray}
W_{H,L}=(\beta T^{1/4})^{\mp 1}(1 + m_{H,L}({\cal K})T^{-3/4} + \cdots),
\end{eqnarray}
\begin{eqnarray}
\nonumber
W_{N}=\beta T^{1/4}\left(1 - \frac {1}{12T} + m_{L}({\cal K})T^{-3/4} +
\cdots\right).
\end{eqnarray}

According to Eqs.(56), at arbitrary changes in $J$ and $p$, with
${\cal K}=$const. retained, the scaling functions are the {\it universal}
functions of the reduced time, $W_{i}(T)$, and the $h_{s}/h_{e},
l_{s}/l_{e}$, and $N/N_{e}$ trajectories are shifted by a factor

$$
p^{\mp1/4}=(J/{\cal K})^{\pm1/6}.
$$

So, the boundaries of the power-law asymptotic regions, the $W_{H}$ and
$|\phi_{i}|$ maxima, the times at which they are attained, $T_{M}$ and
$T_{i}^{m}$, and the $m_{i}$ coefficients are {\it unambiguous} functions of
the parameter ${\cal K}$. Their behavior will be considered in what follows.

\subsubsection{Behavior of $W_{i}({\cal K},T)$ and kinetic diagrams
${\cal K} - T$}

Figs.10,11, and 12 show the plots $W_{H}=p^{1/4}h_{s}/h_{e},
W_{L}=p^{-1/4}l_{s}/l_{e}$, and $W_{N}=p^{-1/4}N/N_{e}$ vs $T=\tau/p$
in the ${\cal K}$ range from ${\cal K}=10^{-11/2}$ to ${\cal K}=10^{13/2}$
(upper panels), and the corresponding ${\cal K}-T$ diagrams of the power-law
asymptotic regions (lower panels), obtained by replotting the data of Fig.5
($J=10^{8}$) for $\tau < \tau_{r}\approx 0.2$ in the corresponding scaling
coordinates. It is seen that with the growing ${\cal K}$ from
${\cal K} \ll 1$, the $W_{i}({\cal K},T)$ trajectories initially shift in
a self-similar manner in accord with the scaling (45)

\begin{eqnarray}
W_{i}({\cal K},T)={\cal K}^{\pm1/6}v_{i}({\cal K}^{2/3}T), \quad
{\cal K}\rightarrow 0,
\end{eqnarray}

(note that for $W_{L}({\cal K},T)$ this scaling takes place only at
$T \gg 1$ whereas at $T \ll {\cal K}^{-2/3}$, according to (28), the scaling
$W_{L}({\cal K},T)=\sqrt{\cal K}u(T)$ takes place) and then at
${\cal K} \gg 1$ they shift in a self-similar manner at "tails" of
$T \ll 1$, in accord with the scaling

\begin{eqnarray}
W_{H,L}({\cal K},T)=\Lambda_{s}({\cal K}T),
\end{eqnarray}
$$
W_{N}({\cal K},T)={\cal K}^{-1/2}\Lambda_{N}({\cal K}T),
$$

converging at ${\cal K}T\rightarrow \infty$ to the universal (${\cal K}$
independent) trajectories (44)

\begin{eqnarray}
W_{i}({\cal K},T)=w_{i}(T), \quad {\cal K}T\rightarrow \infty.
\end{eqnarray}

Fig.11 clearly demonstrates the evolution of the course of the
$W_{L}({\cal K},T)$ trajectories as a result of competition of two opposite
tendencies: (i) growth acceleration of $l_{s}$, associated with the
reflection of $L$ particles from the boundary $x=0$ and (ii) growth
deceleration of $l_{s}$, associated with the establishment of
quasiequilibrium $J^{d}\simeq J$, the former being dominant at any
${\cal K}$ in the vicinity of $T\approx 1$, the latter in the vicinity of
$T\approx T_{*}=1/{\cal K}^{2/3}$ at ${\cal K}<1$ and in the vicinity of
$T\approx T_{q}=1/{\cal K}$ at ${\cal K}>1$, shifting with the growing
${\cal K}$ towards $T\rightarrow 0$.

The general evolution picture of the power-law asymptotic regions with
the growing ${\cal K}$ between the limiting regimes (40) and (41) as well as
of the ${\cal K}$ and $T$ regions, in the confines of which the crossovers
between power-law asymptotics fit the scaling laws (63), (64), and (65),
are demonstrated by the kinetic ${\cal K}-T$ diagrams on the lower panels
of Figs. 10,11, and 12. According to Eqs.(56), the given plots yield an
exhaustive description of evolution of the $h_{s},l_{s}$, and $N$
trajectories at the stage of the $H$ particles diffusion into a semi-infinite
medium at arbitrary $J$ and $p$ changes. In particular, at $p=$ const. the
given plots factually describe the evolution of the $h_{s},l_{s}$, and $N$
trajectories with the growing density of the flux $J$ up to $T=T_{r}$,
where $T_{r}\approx 0.2/p$ is the time, begining with which the reflection
of $H$ particles from the boundary $x=0$ becomes essential.

\subsubsection{Scaling of $h^{M}_{s}$}

From Eqs.(56) it follows that at $p < 1, J \gg J_{c}$, and $\tau < \tau_{r}$
the $p$ and $J$ dependences of the $h_{s}^{M}$ amplitude and the time
$\tau_{M}$, for which $h_{s}$ reaches a maximum, should have the scaling form

\begin{eqnarray}
h_{s}^{M}=h_{e}p^{-1/4}M({\cal K}), \quad \tau_{M}=pT_{M}({\cal K}).
\end{eqnarray}

In the limit of small ${\cal K} \ll 1$, according to (63),(46), and (49),
the scaling $M({\cal K})$ and $T_{M}({\cal K})$ functions have the
asymptotics

\begin{eqnarray}
M({\cal K})=0.7221{\cal K}^{1/6}, \quad T_{M}({\cal K})=1.125{\cal K}^{-2/3}.
\end{eqnarray}

In the opposite limit of large ${\cal K} \gg 1$ the quantity $M({\cal K})$
reaches its limiting value $M({\cal K}\rightarrow \infty)= 1$.
By differentiating (60), in this limit we find

\begin{eqnarray}
M({\cal K})=1- \left(\frac {\ln16{\cal K}}{8\pi{\cal K}}\right)^{1/2}
\left(1 + \frac {1}{\ln16{\cal K}}\right),
\end{eqnarray}
\begin{eqnarray}
\nonumber
T_{M}({\cal K})= \frac {2}{\ln16{\cal K}},
\end{eqnarray}

whence it follows that $M({\cal K})$ approaches $1$ as $1-M\propto
\sqrt {\ln {\cal K}/{\cal K}}$, and a rapid drop of $T_{M}({\cal K})$
tranforms at ${\cal K} \gg 1$ to a slow logarithmic one.
In Fig.12 are presented the numerical data of Fig.6 replotted in the
scaling coordinates $p^{1/4}h_{s}^{M}/h_{e}$ - ${\cal K}$ and
$\tau_{M}/p$-${\cal K}$, respectively. It is seen, that in complete
agreement with (62), the points, calculated for different $p$ and $J$
values, fit the scaling functions $M({\cal K})$ and $T_{M}({\cal K})$ which
at small ${\cal K} \ll 1$ and large ${\cal K} \gg 1$ approach, respectively,
asymptotics (67) and (68), shown in dashes.

\subsubsection{Scaling of $\phi^{m}_{i}$ and $m_{i}$}

In Figs.13a and b are drawn, respectively, the $|\phi_{i}|^{m}$ and
$T^{m}_{i}$ vs ${\cal K}$ dependences derived from the data of Figs.8 and 9,
and the analogous data for some other $p < p_{s}$ and $J > J_{1/4}$ values.
The points, calculated in a broad range of $p$ and $J$, are seen to fit, in
accord with (56), the scaling functions $\phi^{m}({\cal K})$ and
$T^{m}_{i}({\cal K})$, which in the limit of small ${\cal K} \ll 1$ approach
the asymptotics (55) (dashed lines)

$$
\phi^{m}_{H}=0.00108, \quad T^{m}_{H}= 17.3{\cal K}^{-2/3},
$$

and in the limit of large ${\cal K}\gg 1$ merge and approach the
asymptotics (52) (dashed lines)

$$
|\phi_{i}|^{m}=0.00798, \quad T^{m}_{i}=2.51.
$$

According to Eqs.(33),(50), and (56), at $p \ll p_{s}$ and $J \gg J_{1/4}$
in the range of $T^{m}_{H} \ll T < T_{r}$ the $\tilde {\phi}_{i}({\cal K},T)$
functions ought to asymptotically decay by the law

$$
\tilde {\phi}_{i}({\cal K},T)=m_{i}({\cal K})/T^{3/4}+\cdots
$$

with the coefficients $m_{i}({\cal K})$, satisfying the condition

$$
m_{H}({\cal K}) + m_{L}({\cal K})= -\beta/4\sqrt {{\cal K}}.
$$

In Fig.13c are drawn the ${\cal K}$ dependences of $m_{H}$ and $|m_{L}|$,
derived from the data of Figs. 8 and 9, and the analogous data for some other
$p \ll p_{s}$ and $J \gg J_{1/4}$. The points, calculated in a broad range
of $p$ and $J$, are seen to fit the scailing functions $|m_{i}({\cal K})|$,
which in the limit of small ${\cal K} \ll 1$ approach the asymptotics (54),
(dashed lines)

$$
|m_{i}({\cal K})|=m^{(0)}_{i}/\sqrt {{\cal K}},
$$

and in the limit of large ${\cal K} \gg 1$ approach the asymptotics (51)
(dashed lines)

$$
m^{\infty}_{H}=-m^{\infty}_{L}=0.023.
$$

Figs.(7), (8), (9), and (14) give a complete picture of the crossovers to
the universal $(\beta\tau)^{\mp 1/4}$ asymptotics in a broad range of
$0 < {\cal K} < \infty$, the main features of which can be summarized as
follows:

1) With growing ${\cal K}$ the $\phi_{H}^{m}$ amplitude grows from
the ${\cal K}$-independent asymptotics (55) at ${\cal K} \approx 10^{-3}$ to
the ${\cal K}$-independent asymptotics (52) at ${\cal K} > 10^{5}$. The time
of its attaining, $T_{H}^{m}$, drops up to ${\cal K} \approx 1$ according to
the $p$-independent asymptotics (55), reaching the $I$-independent
asymptotics (52) in the same range of ${\cal K} > 10^{5}$.

2) With growing ${\cal K}$ at the critical point ${\cal K}_{m}=3.0$
the $|\phi_{L}^{m}|$ maximum appears, and its amplitude then drops to the
${\cal K}$-independent asymptotics (52) at ${\cal K} > 10^{5}$. The time
of its attaining, $T_{L}^{m}$, grows and approaches the $I$-independent
asymptotics (52) in the same range of ${\cal K} > 10^{5}$. In the second
critical point ${\cal K}_{c}=7.3$ the ${\cal K} > {\cal K}_{c}$ region
appears, within which the $\phi_{L}$ curves cross zero, changing the sign
herewith.

3) Surprisingly, in the narrow vicinity of the critical point ${\cal K}_{m}$
both the time $T_{H}^{m}$ and the $|m_{L}|$ coefficient approach the
$p$-independent asymptotics (55) and (54), respectively, whereas the $m_{H}$
coefficient approaches the $I$-independent asymptotics (51): the formers
change within the ${\cal K} < {\cal K}_{m}$ range as ${\cal K}^{-2/3}$ and
${\cal K}^{-1/2}$, respectively, the latter remains constant within the
${\cal K} > {\cal K}_{m}$ range.

\section{\bf LONG-TIME RELAXATION DYNAMICS ($\omega\tau > 1$).}

In the light of the results of the previous Section IV, we return, in
conclusion, to the analysis of the long-time relaxation dynamics of
$h_{s},l_{s}$, and $N$, which, according to Section III can be given as

\begin{eqnarray}
\nonumber
h_{s}=h_{e}(1-\sigma_{H}),
\end{eqnarray}
\begin{eqnarray}
l_{s}=l_{e}(1-\sigma_{L}),
\end{eqnarray}
\begin{eqnarray}
\nonumber
N=N_{e}(1-\sigma_{N}),
\end{eqnarray}

where $\sigma_{H}=\tilde {h}_{s}/h_{e},\sigma_{L}=\tilde {l}_{s}/l_{e}$, and
$\sigma_{N}= \tilde {n}/N_{e}$ at sufficiently large $\omega\tau > 1$ decay
by the law

\begin{eqnarray}
\nonumber
\sigma_{i}=A_{i}e^{-\omega\tau}(1+O(e^{-\omega\tau})),
\end{eqnarray}

and the coefficients $A_{i}$ are related as

\begin{eqnarray}
A_{H}=r_{s}A_{L}=A_{N}\sqrt {\omega}\cot \sqrt {\omega}.
\end{eqnarray}

By preserving in (8) the next-to-leading terms

\begin{eqnarray}
\nonumber
\tilde {h}_{s}= {\cal A}_{H}\cos(\sqrt {\omega}x)e^{-\omega\tau} +
{\cal B}_{H}\cos(\sqrt {2\omega}x)e^{-2\omega\tau}+ ...,
\end{eqnarray}
\begin{eqnarray}
\nonumber
\tilde {l}_{s}= {\cal A}_{L}\cos(\sqrt {p\omega}x)e^{-\omega\tau} +
{\cal B}_{L}\cos(\sqrt {2p\omega}xe^{-2\omega\tau} + ...,
\end{eqnarray}

from Eq.(7) we easily find

\begin{eqnarray}
\sigma_{i}=A_{i}e^{-\omega\tau}(1 + B_{i}\sigma_{N}+...),
\end{eqnarray}

where the coefficients $B_{i}$ are defind from the relations

\begin{eqnarray}
\nonumber
B_{H}= \sqrt {p\omega}\cot \sqrt {p\omega}\cot \sqrt {2\omega}/Q,
\end{eqnarray}
\begin{eqnarray}
\nonumber
B_{L}= \sqrt {p\omega}\cot \sqrt {\omega}\cot \sqrt {2p\omega}/Q,
\end{eqnarray}
\begin{eqnarray}
\nonumber
B_{N}= \sqrt {p\omega}\cot \sqrt {p\omega}\cot \sqrt {\omega}/\sqrt {2}Q,
\end{eqnarray}
\begin{eqnarray}
\nonumber
Q= \cot \sqrt {2\omega} + \sqrt {p}\cot \sqrt {2p\omega} - \sqrt {2\omega/J}.
\end{eqnarray}

By substituting now (69) into (14) and (15) and introducing the designation
${\cal D}=\sigma_{H}\sigma_{L}$, after having excluded $\hat\sigma_{H}(s)$
and $\hat\sigma_{N}(s)$, we find

\begin{eqnarray}
\hat \sigma_{L}(s)= \frac {\coth\sqrt {s} -
(1-s\hat {\cal D}(s))\sqrt {p}\coth \sqrt {ps}+ \sqrt {s/J}}
{s(\coth \sqrt {s}+\sqrt {p}\coth \sqrt {ps}+\sqrt {s/J})}
\end{eqnarray}

In accord with the results of Section III, the main pole of Eq.(72)
($s=-\omega$) defines the relaxation rate $\omega$ as the least positive
root of (9) and yields the relaxation "amplitude" $A_{L}$ in the form

\begin{eqnarray}
A_{L}=A_{L0}(1+\case 1/2 \omega \hat {\cal D}(-\omega))
\end{eqnarray}

where

\begin{eqnarray}
A_{L0}= \frac {4\sqrt {p/\omega}\cot \sqrt {p\omega}}
{\sin^{-2}\sqrt
{\omega} + p\sin^{-2}\sqrt {p\omega}+ 1/\sqrt {J}}
\end{eqnarray}

and $\hat {\cal D}(-\omega)=\int \limits_{0}^{\infty}\sigma_{H}\sigma_{L}
e^{\omega\tau}d\tau$.
In the diffusion-controlled limit of our concern here, $J\rightarrow \infty$,
the coefficients $B_{i}$ change from $B_{N}=B_{L}=B_{H}-1=0.1079... $ at
$p\rightarrow 0$ to $B_{i}\rightarrow 0$ at $p\rightarrow 1$, therefore
given that the next after the main root of (9) is an order of magnitude
above, we conclude that down to $\tau\sim \omega^{-1}$ the asymptotic
behavior of $\sigma_{L}$ has the form $\sigma_{L}=A_{L}e^{-\omega\tau}
(1+B_{L}\sigma_{N}+ ...)\simeq A_{L}e^{-\omega\tau}$. By matching this
asymptotics at $\tau=\omega^{-1}$ with that of (32), which for $p \ll 1$
assumes the form
$\sigma_{L}=(1-\beta\tau^{1/4}(1+\phi_{L}))\simeq 1-\beta\tau^{1/4}$, in
the limit $p\rightarrow 0$ we find

\begin{eqnarray}
A_{L}\simeq e(1-\beta/\omega^{1/4})\simeq 0.5
\end{eqnarray}

From (73) and (75) it follows that at $p\rightarrow 0$ the quantity
$\hat {\cal D}(-\omega)\sim -0.1$ and the difference of $A_{L}$ from
$A_{L0}$ does not exceed $20\%$. Obviously, with the growing $p$ this
difference can only go down, therefore by assuming in the first
approximation $A_{L}\simeq A_{L0}$ and extrapolating the
${\cal D}\simeq -A_{L0}^{2}e^{-2\omega\tau}$ asymptotics down to $\tau=0$
($\hat {\cal D}(-\omega) \simeq -A_{L0}^{2}/\omega$), we finally come to
simple expressions

\begin{eqnarray}
A_{L}= A_{H}/r_{s} \sim A_{L0}(1-\case 1/2 A_{L0}^{2}),
\end{eqnarray}
\begin{eqnarray}
A_{N}= \frac {4(1-\case 1/2 A_{L0}^{2})}
{\omega(\sin^{-2} \sqrt {\omega} + p\sin^{-2} \sqrt {p\omega})},
\end{eqnarray}

which, according to the numerical data, differ from the exact values
by no more than $1-2\%$ for any $p$. Fig.15 gives, as an illustration,
the time dependences $|\sigma_{H}(\tau)|$ and $\sigma_{L}(\tau)$,
calculated numerically at $J=10^{6}$ for $p=10^{-4},0.5$ and at $J=10^{8}$
for $p=0.9$ and compared with the results of the calculation from the
expressions $\sigma_{i}=A_{i}e^{-\omega\tau}$, (9), (76) and (74).
The excellent agreement between the numerical and analytical data
is evident.

\begin{figure}
\caption{Schematic illustration of the processes of dissociation,
desorption, surface migration and diffusion into the bulk in the
system $A + B \rightarrow 0$ with input of $A$ and $B$ particles.
$\ell_{A}^{D}=\sqrt{D_{A}t}$ and $\ell_{B}^{D}=\sqrt{D_{B}t}$ are the
diffusion lengths of $A's$ and $B's$ particles, respectively.}
\label{fig1}
\end{figure}

\begin{figure}
\caption{Time dependences $h_{s}(\tau)/h_{e}$, calculated numerically at
flux densities $J=1, J_{c}=7.250031, 10^{2},
10^{3}$, and $10^{4}$ (bottom to top) for $p=0.1$.}
\label{fig2}

\end{figure}

\begin{figure}
\caption{Dependence of $r_{s}$ on $\sqrt {J/J_{c}}$, calculated from
Eq.(5) for $p=0.1$. Left and right insets demonstrate the specific
features of relaxation of surface concentrations at small $J \ll J_{c}$
and large $J \gg J_{c}$, respectively.}
\label{fig3}
\end{figure}

\begin{figure}
\caption{Time dependences $h_{s}(\tau)/h_{e}$ and $l_{s}(\tau)/l_{e}$,
calculated numerically at $J=J_{c}$ (open circles) and
$J=10^{3}$ (filled circles) for $p=0.1$.}
\label{fig4}
\end{figure}

\begin{figure}
\caption{(a) Time dependences $h_{s}(\tau)/h_{e}, l_{s}(\tau)/l_{e}$, and
$N(\tau)/N_{e}$, calculated numerically at $p=1, 10^{-1}, 10^{-2}, 10^{-3},
10^{-4}, 10^{-5}, 10^{-6}, 10^{-7}, 10^{-8}$, and $10^{-9}$ for
$J=10^{8}$ (the arrows point to the displacement directions of the
corresponding trajectories at a change in $p$ from $p=1$ to $p=10^{-9}$).
The trajectories for $p=1$ and $p=10^{-9}$ are given in bold lines;
(b), (c) Kinetic $p - \tau$ diagrams of the regions of power-law asymptotics
and of steady state (shaded by light gray) for $h_{s}(\tau)/h_{e}$ (b) and
$l_{s}(\tau)/l_{e}$ (c) trajectories calculated numerically at $J=10^{8}$
according to condition (48). The characteristic times $\tau_{H}=1,
\tau_{L}=p, \tau_{q}=1/J\sqrt {p}$, and $\tau_{*}=1/J^{2/3}$ are shown in
dashed lines. The intersection point of $\tau_{L}, \tau_{q}$, and
$\tau_{*}$ ($\tau_{q}=\tau_{L}=\tau_{*}=4.64158\times 10^{-6})$ is marked
off by filled circle. The arrow shows the direction of displacement of
this point along the $\tau_{L}$ line with the growing flux density $J$.
In Fig.5c the $\beta\tau^{1/4}$ asymptotic region for the $N(\tau)/N_{e}$
trajectories is distinquished by dark gray.}
\label{fig5}
\end{figure}

\begin{figure}
\caption{Dependences of $h_{s}^{M}/h_{e}$ (a) and $\tau_{M}$ (b)
on $J$, calculated numerically at $p=10^{-1}, 10^{-2}, 10^{-3}, 10^{-4},
10^{-5}$, and $10^{-6}$ (bottom to top and top to bottom in (a) and (b),
respectively). Dashed lines show the asymptotics (49).}
\label{fig6}
\end{figure}

\begin{figure}
\caption {Time dependences $\phi_{H}(\tau)$ (a) and $\phi_{L}(\tau)$ (b),
calculated numerically at $p=10^{-4}$ for $J=10^{4}, 10^{5}, 10^{6}, 10^{7},
10^{8}, 10^{9}, 10^{10}, 10^{11}$, and $10^{12}$ (right to left).}
\label{fig7}
\end{figure}

\begin{figure}
\caption{Data of Fig.7 for sections $\phi_{H}>0$ (a) and $\phi_{L}<0$ (b)
replotting in double logarithmic scale. Dashed lines show the slope of
$-3/4$. Bold lines show the $\phi^{0}$ curve, calculated from Eq.(39).}
\label{fig8}
\end{figure}

\begin{figure}
\caption{Time dependences $\phi_{H}(\tau)$ (a) and $|\phi_{L}(\tau)|$ (b)
for sections $\phi_{H} > 0$ and $\phi_{L} < 0$, respectively, calculated
numerically at $J=10^{6}$ for $p=10^{-2n/3}$ with $n=3, 4, 5, 6, 7, 8$, and
$9$(right to left). Dashed lines show the slope of $-3/4$. Bold lines show
the $\phi^{0}$ curve, calculated from Eq.(39).}
\label{fig9}
 \end{figure}

\begin{figure}
\caption{(a) Scaling functions $W_{H}({\cal K},T)$, derived from the data
of Fig5a ($p$ from $p=10^{-1}$ to $10^{-9}$, $\tau \leq 0.2$) by replotting
in scaling coordinates $p^{1/4}h_{s}/h_{e}$ vs $T=\tau/p$.
(b) Kinetic ${\cal K} - T$ diagram of the regions of power-law asymptotics
of $W_{H}({\cal K},T)$, derived from the data of Fig.5b ($\tau \leq 0.2$)
by replotting in scaling coordinates ${\cal K}=p^{3/2}J$ vs $T=\tau/p$.
Note the different scale on the $T$ axis in (a) and (b).}
\label{fig10}
\end{figure}

\begin{figure}
\caption{(a) Scaling functions $W_{L}({\cal K},T)$, derived from the data
of Fig5a ($p$ from $p=10^{-1}$ to $10^{-9}$, $\tau \leq 0.2$) by replotting
in scaling coordinates $p^{-1/4}l_{s}/l_{e}$ vs $T=\tau/p$.
(b) Kinetic ${\cal K} - T$ diagram of the regions of power-law asymptotics
of $W_{L}({\cal K},T)$, derived from the data of Fig.5c ($\tau \leq 0.2$)
by replotting in scaling coordinates ${\cal K}=p^{3/2}J$ vs $T=\tau/p$.
Note the different scale on the $T$ axis in (a) and (b).}
\label{fig11}
\end{figure}

\begin{figure}
\caption{(a) Scaling functions $W_{N}({\cal K},T)$, derived from the data
of Fig5a ($p$ from $p=10^{-1}$ to $10^{-9}$, $\tau \leq 0.2$) by replotting
in scaling coordinates $p^{-1/4}N/N_{e}$ vs $T=\tau/p$.
(b) Kinetic ${\cal K} - T$ diagram of the regions of power-law asymptotics
of $W_{N}({\cal K},T)$, derived from the data of Fig.5c ($\tau \leq 0.2$)
by replotting in scaling coordinates ${\cal K}=p^{3/2}J$ vs $T=\tau/p$.
As contrasted from Fig.5c, the regions with $n=1/2$ and $n=1$ are added.
Note the different scale on the $T$ axis in (a) and (b).}
\label{fig12}
\end{figure}

\begin{figure}
\caption{(a) Collapse of the dependences of $h_{s}^{M}/h_{e}$
on $J$ from Fig.6a to the scaling function $M({\cal K})$ in scaling
coordinates $p^{1/4}h_{s}^{M}/h_{e}$ vs ${\cal K}=p^{3/2}J$.
Dashed lines show the asymptotics (67) and (68). (
b) Collapse of the dependences of $\tau_{M}$ on $J$ from Fig.6b to the
scaling function $T_{M}({\cal K})$ in scaling coordinates
$\tau_{M}/p$ vs ${\cal K}=p^{3/2}J$. Dashed lines show the asymptotics
(67) and (68).}
\label{fig13}
\end{figure}

\begin{figure}
\caption{Collapse of calculated numerically for $p=10^{-4}$ (circles) and
$p=10^{-5}$ (diamonds) dependences $|\phi_{i}^{m}(J)|, T^{m}_{i}(J)$, and
$|m_{i}(J)|$, and calculated numerically for $J=10^{6}$ (stars) and
$J=10^{7}$ (squares) dependences $|\phi^{m}_{i}(p)|, T^{m}_{i}(p)$, and
$|m_{i}(p)|$ to scaling functions $|\phi^{m}_{i}({\cal K})|$ (a),
$T^{m}_{i}({\cal K})$ (b), and $|m_{i}({\cal K})|$ (c), respectively
Filled symbols stand for $i=H$, open symbols stand for $i=L$. In Fig.14a,b
filled circle $m$ marks off the point ${\cal K}_{m}=3.0$ with which the
${\cal K} \geq {\cal K}_{m}$ region begins where a maximum on the
$|\phi_{L}|$ curves appears. Open circle $c$ marks off the point
${\cal K}_{c}=7.3$ with which the region ${\cal K} \geq {\cal K}_{c}$
of crossing of the $\phi_{L}$ curve with zero begins. Dashed lines
show the asymptotics (52) and (55)(a),(b), and (54) and (51)(c).}
\label{fig14}
\end{figure}

\begin{figure}
\caption{Time dependences $|\sigma_{H}(\tau)|$ (filled circles) and
$\sigma_{L}(\tau)$ (open circles), calculated numerically at $p=10^{-4}$ and
$p=0.5$ for $J=10^{6}$ and at $p=0.9$ for $J=10^{8}$ (for the sake of
clarity, the data are given beginning with $\tau=10^{-2}$.
Straight bold lines show $|\sigma_{H}(\tau)|$ and $\sigma_{L}(\tau)$,
calculated from the equations $\sigma_{i}=A_{i}e^{-\omega\tau}$, (9), (76),
and (74).}
\label{fig15}
\end{figure}

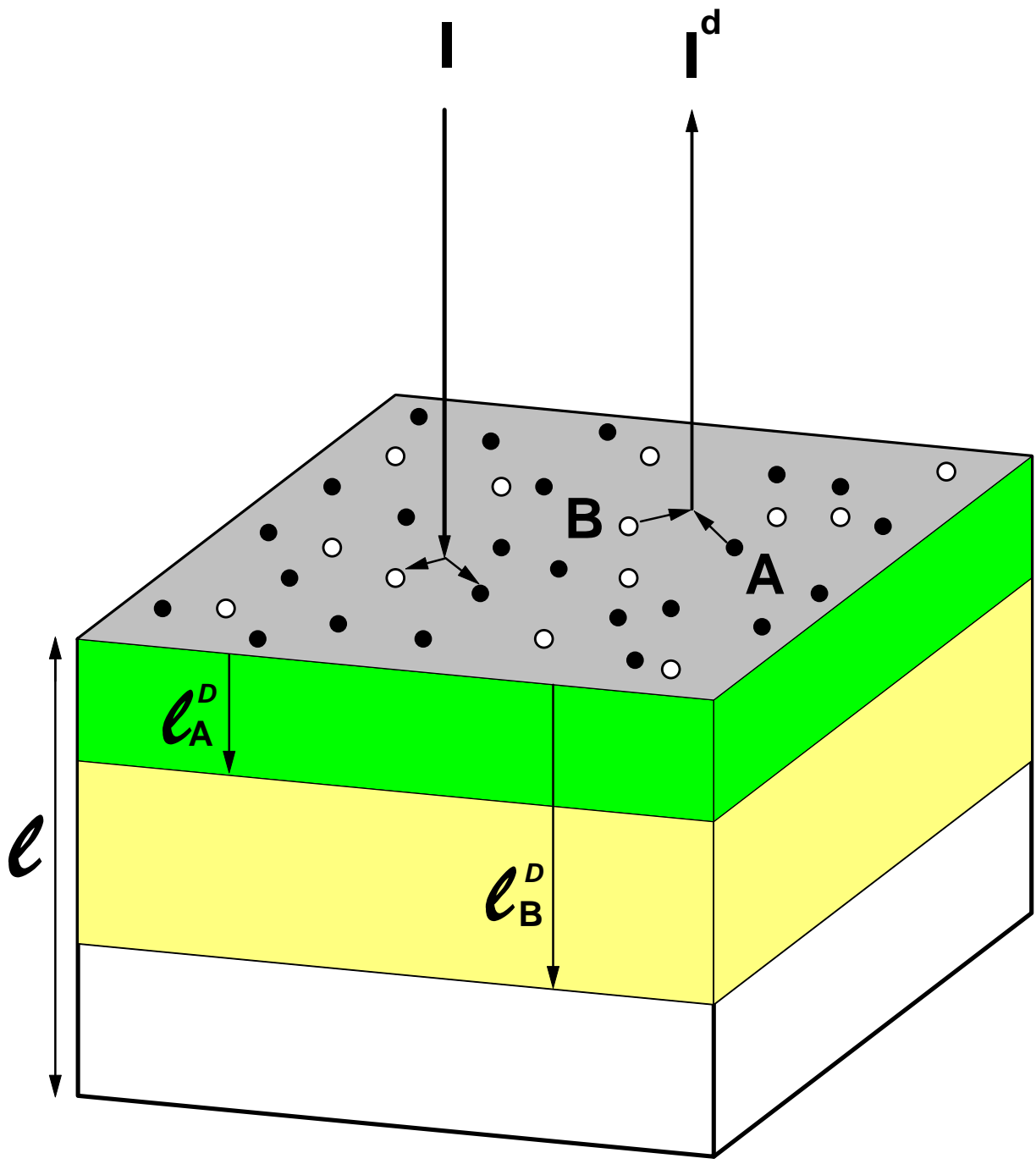
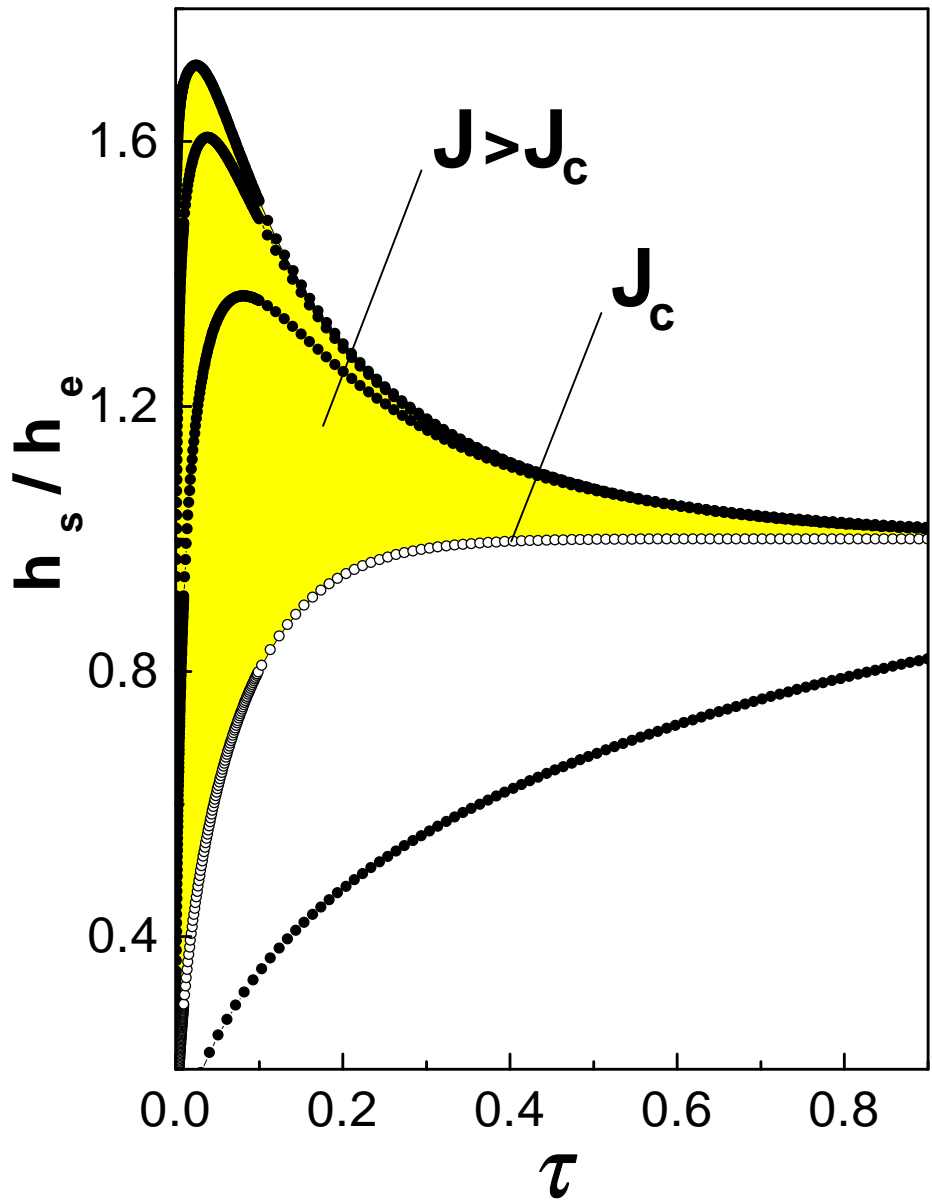
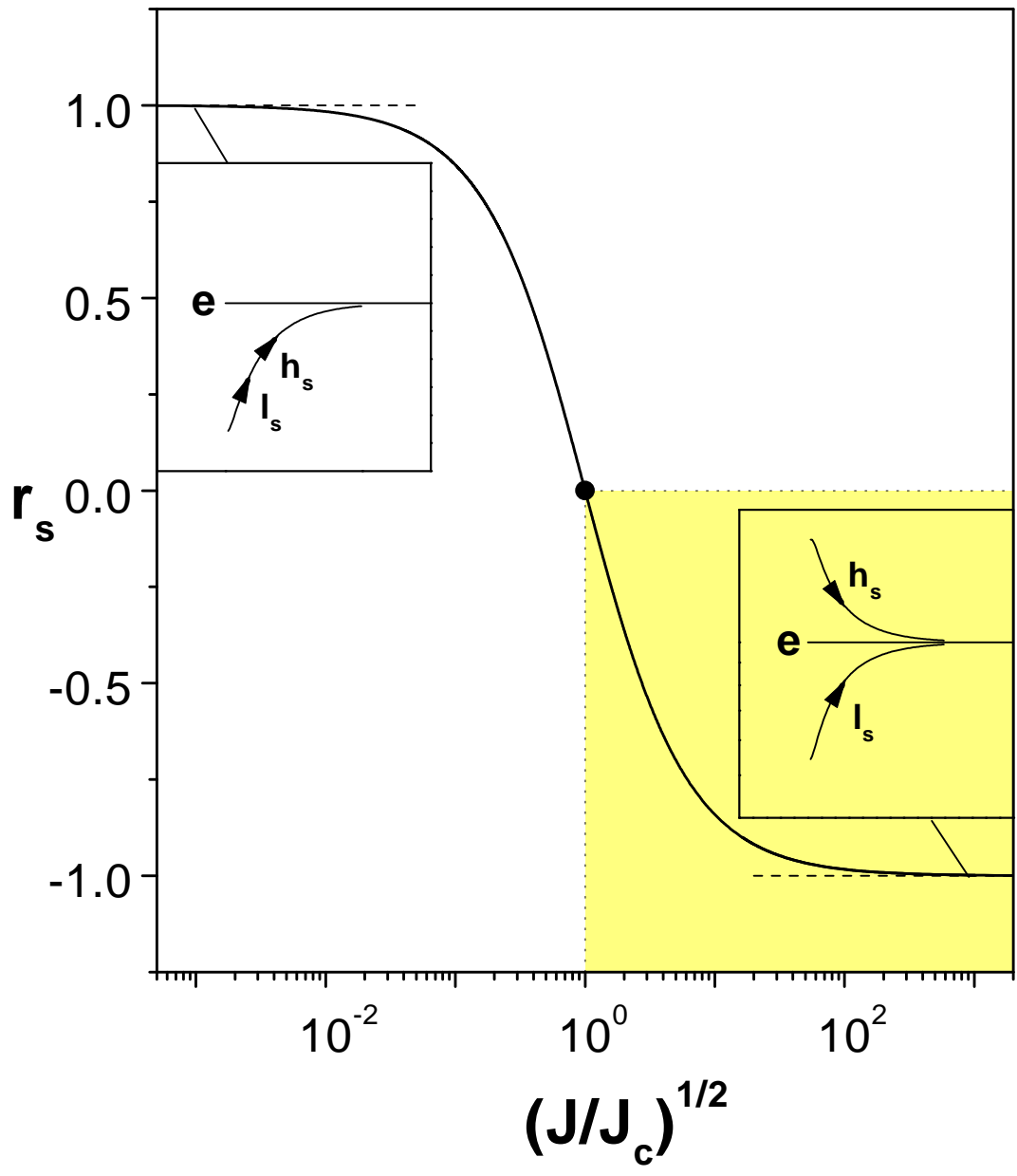
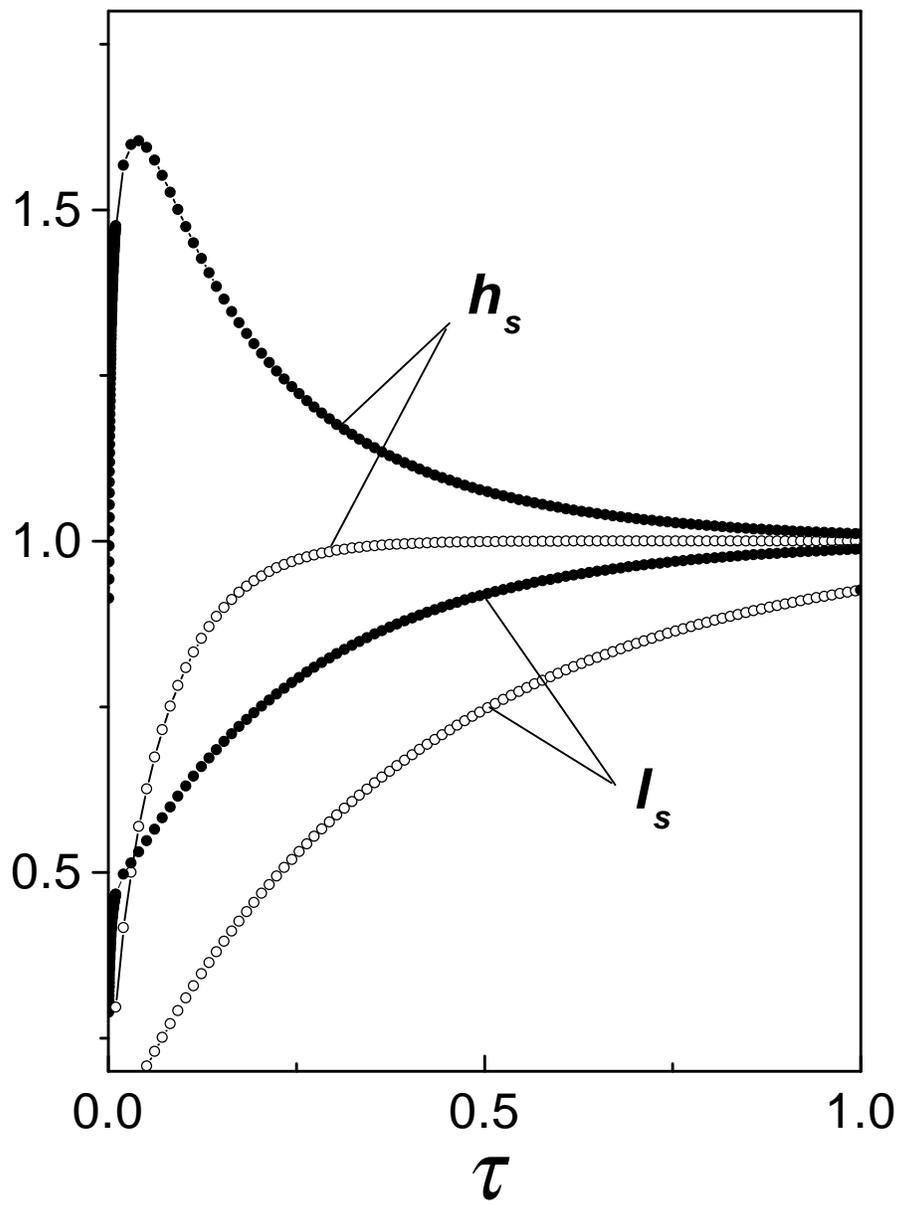
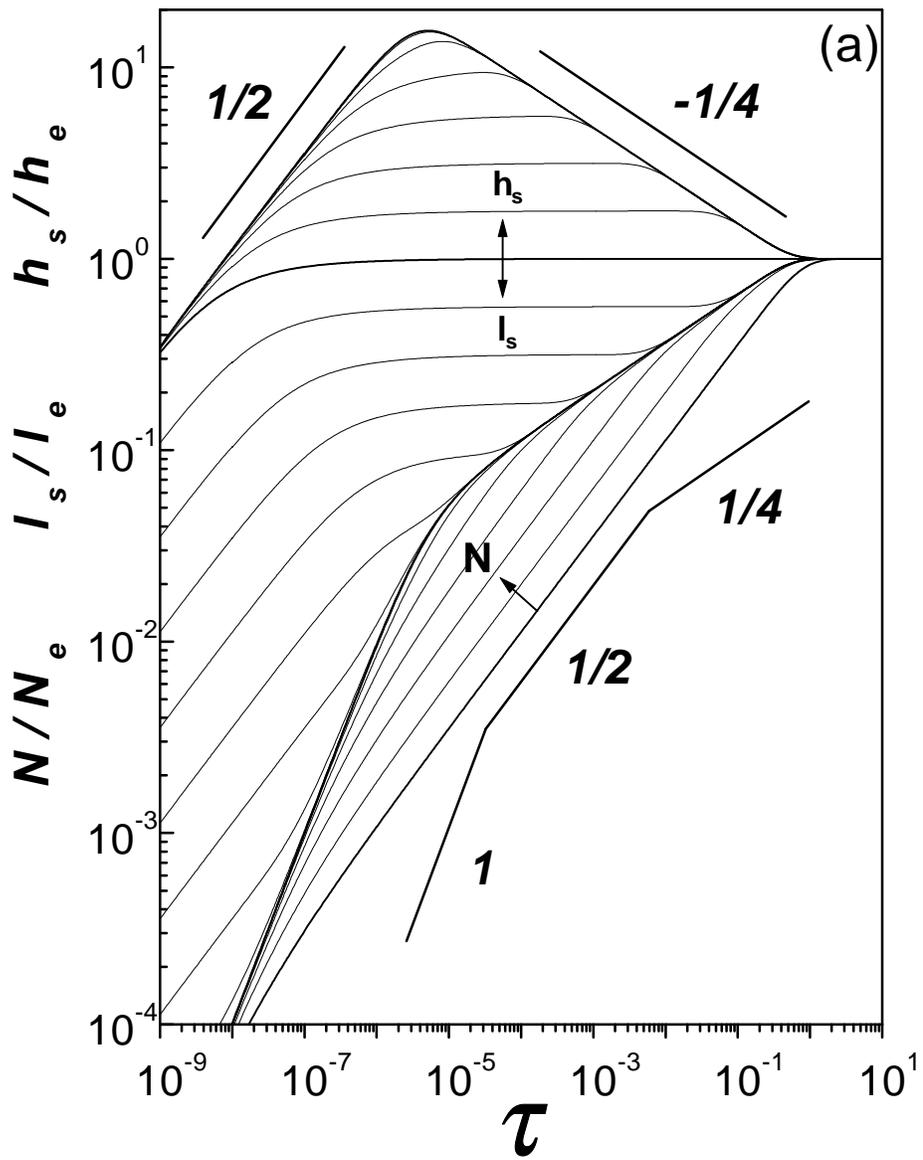
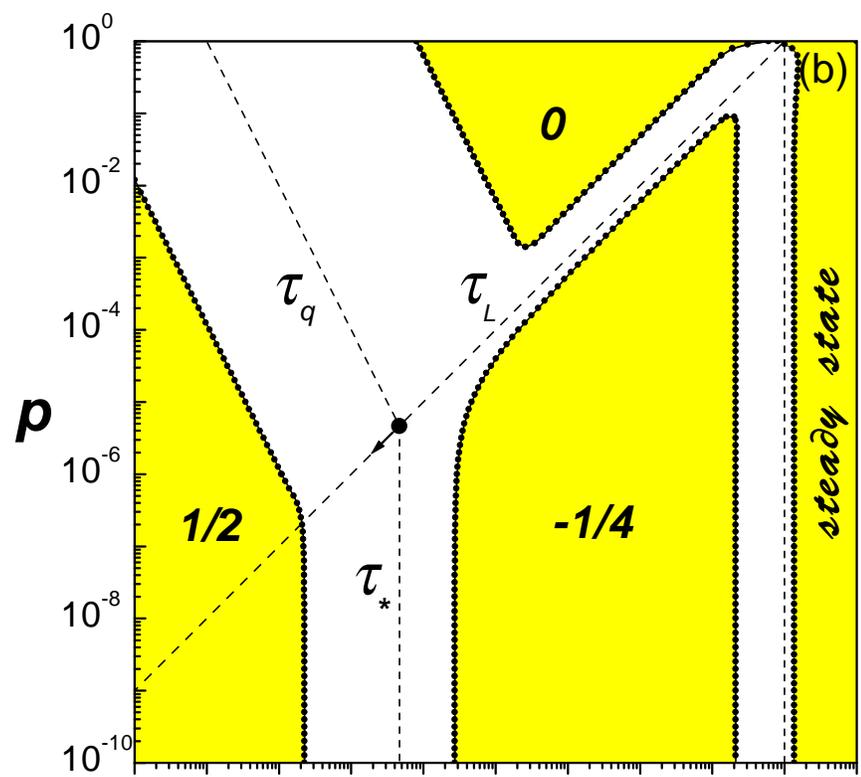
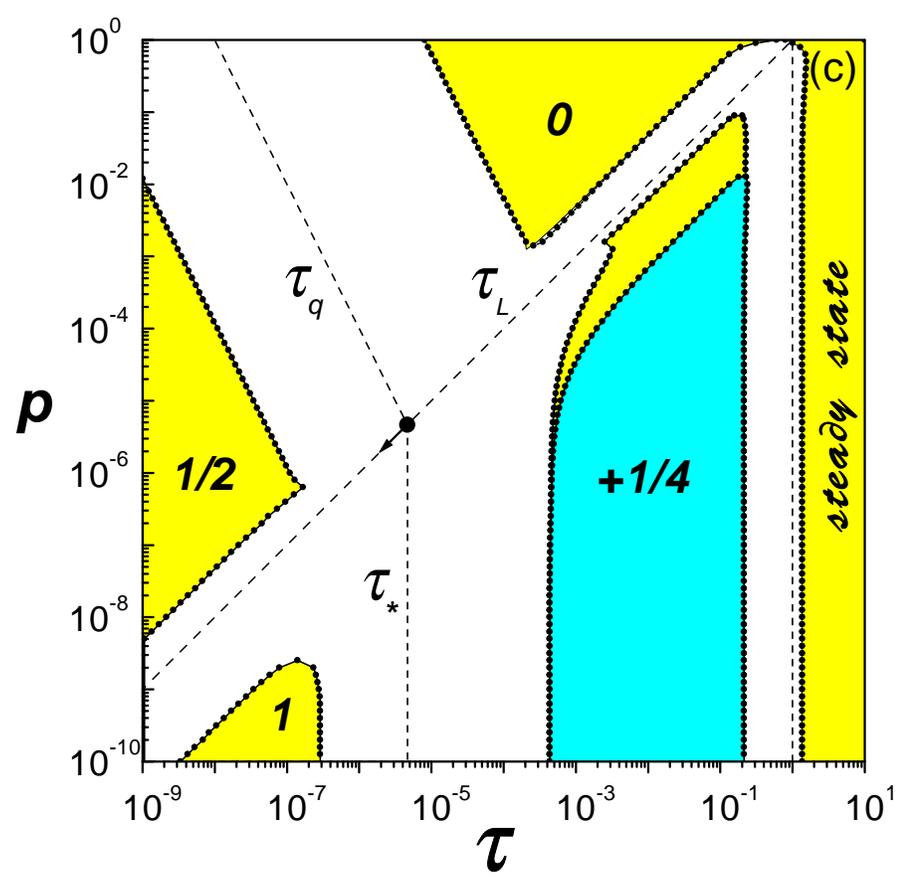
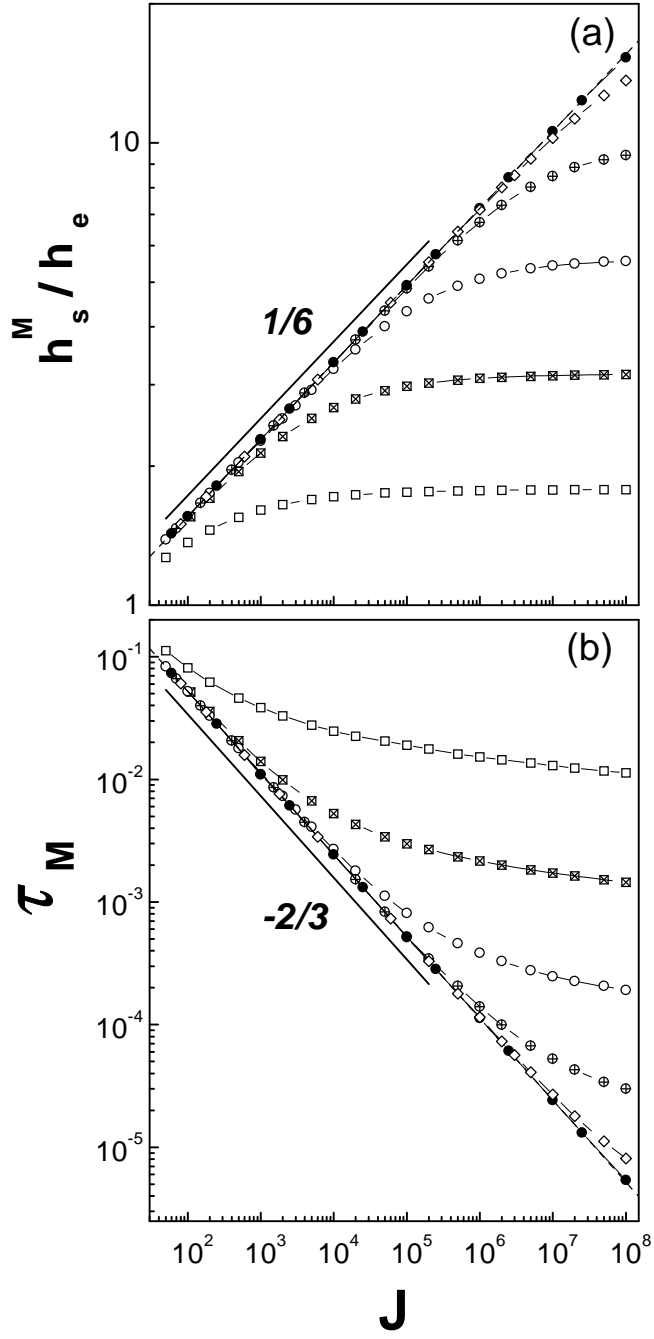
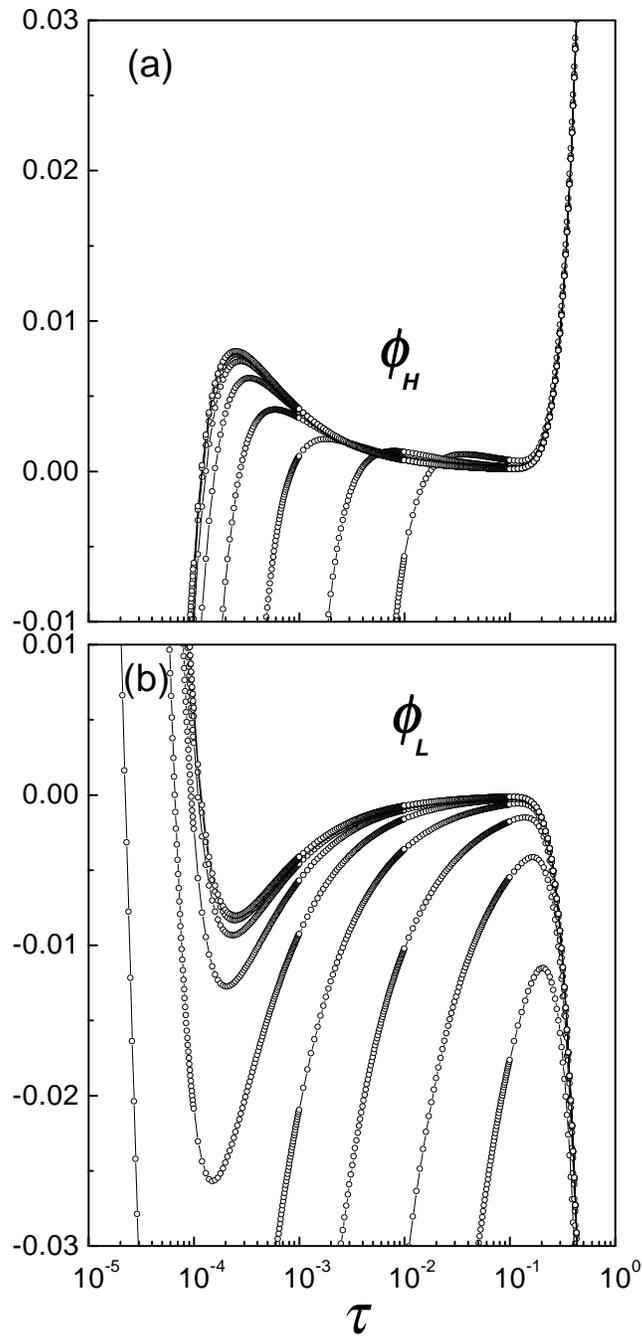
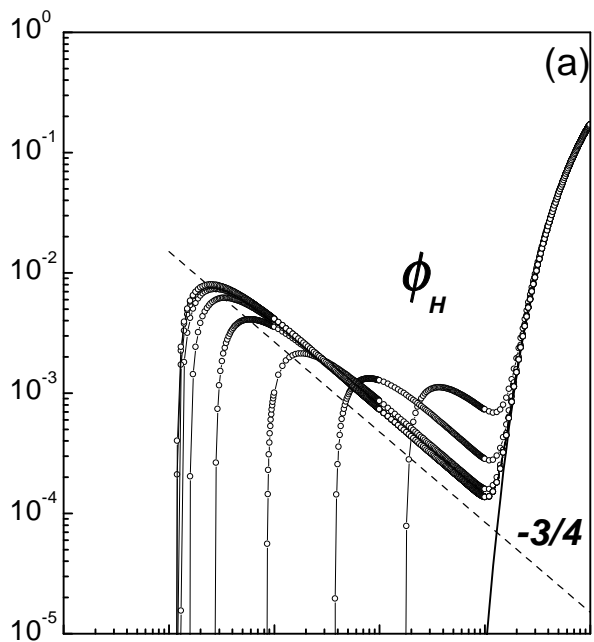
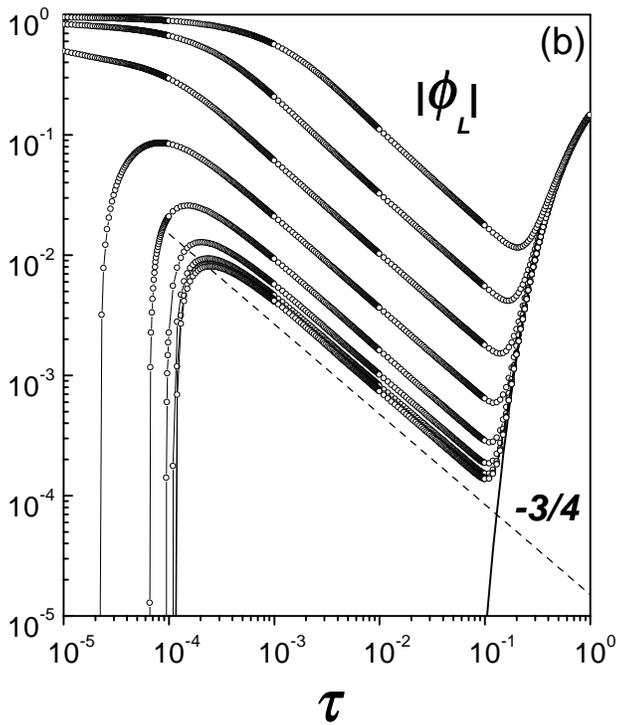
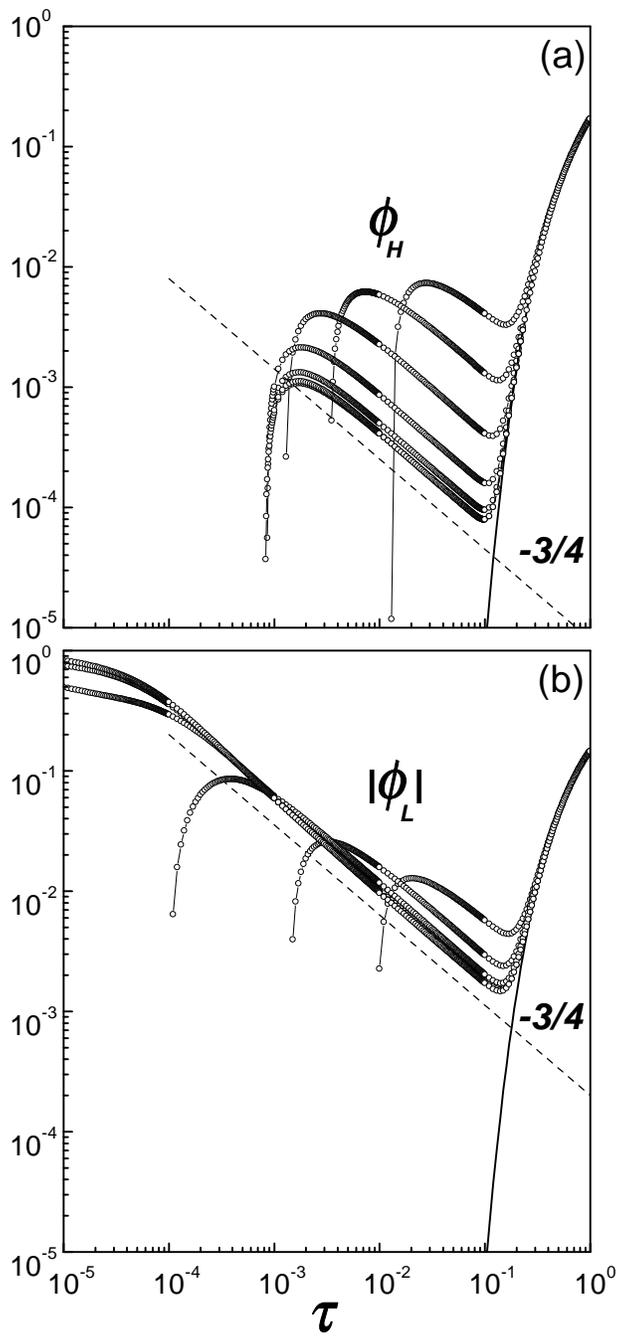
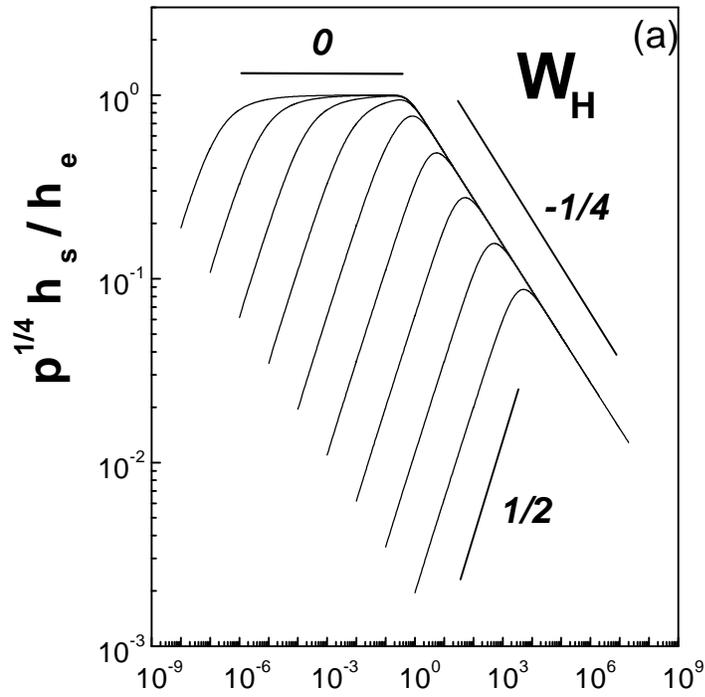
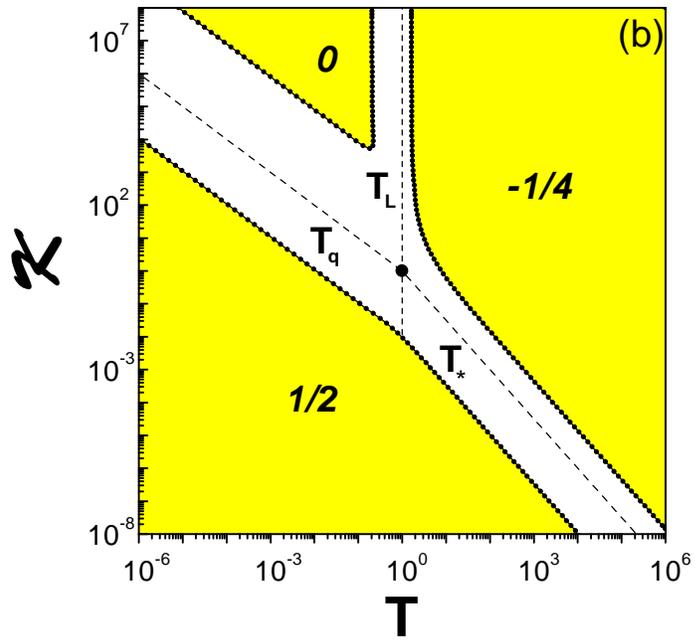
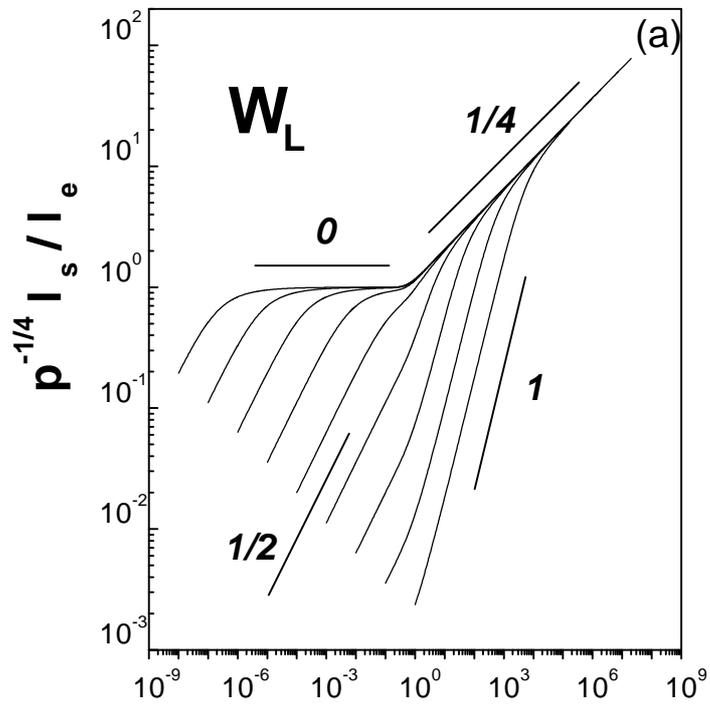
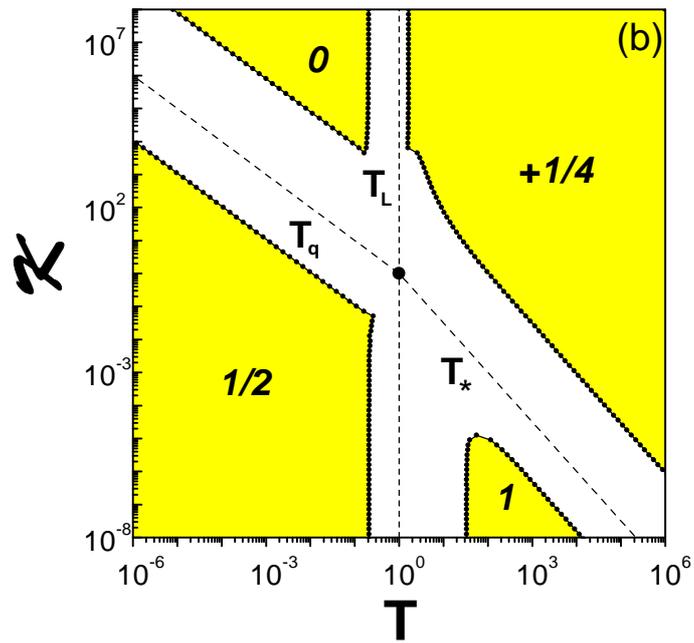
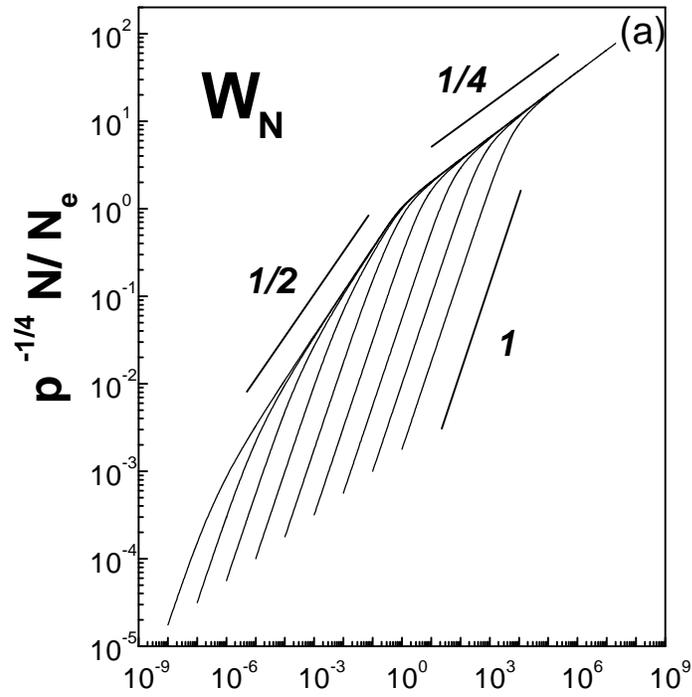
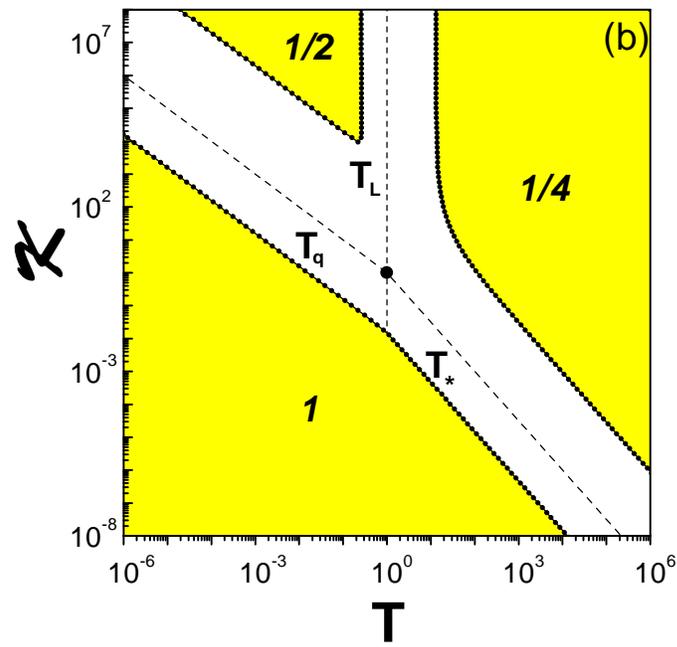
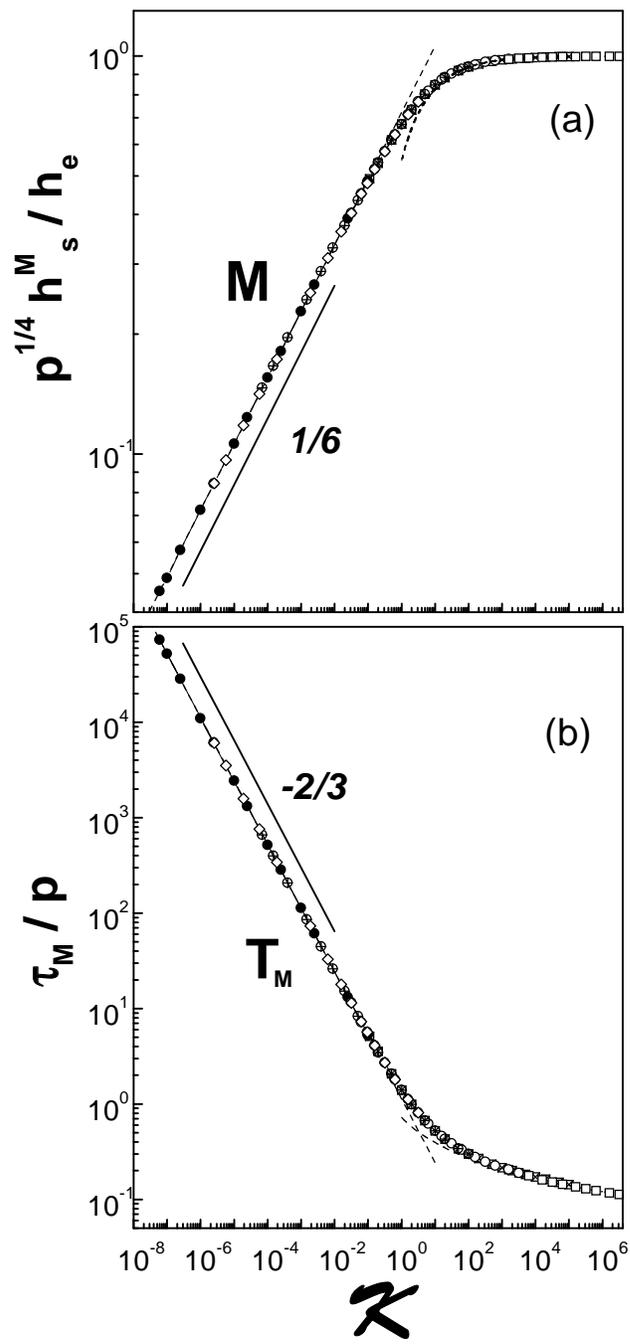
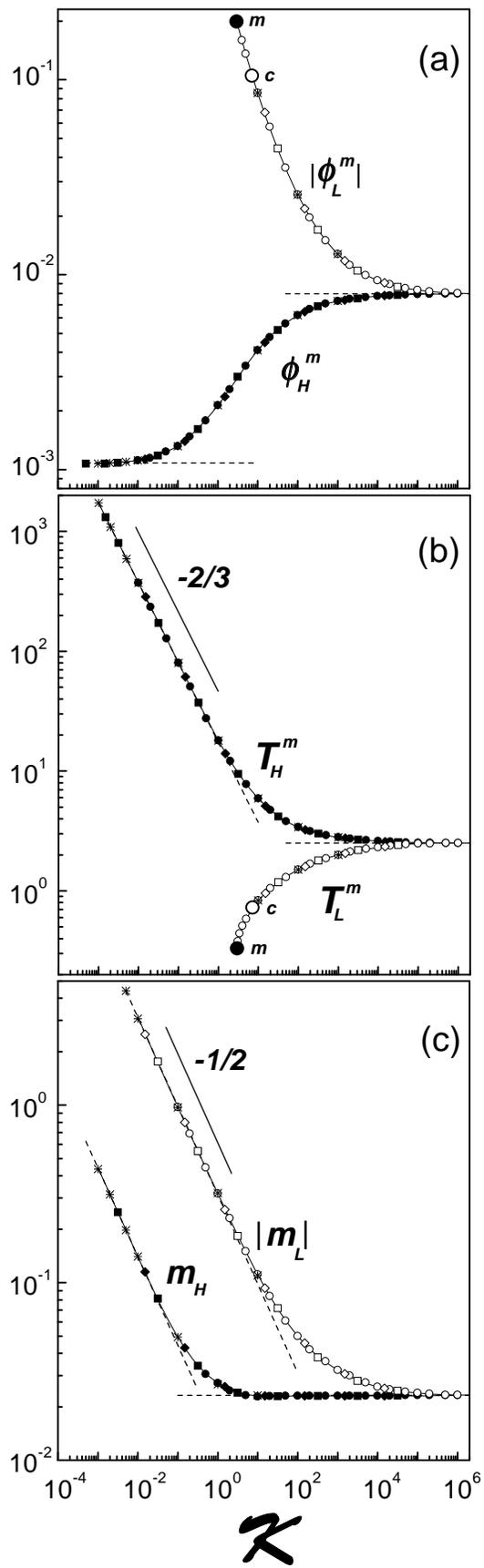
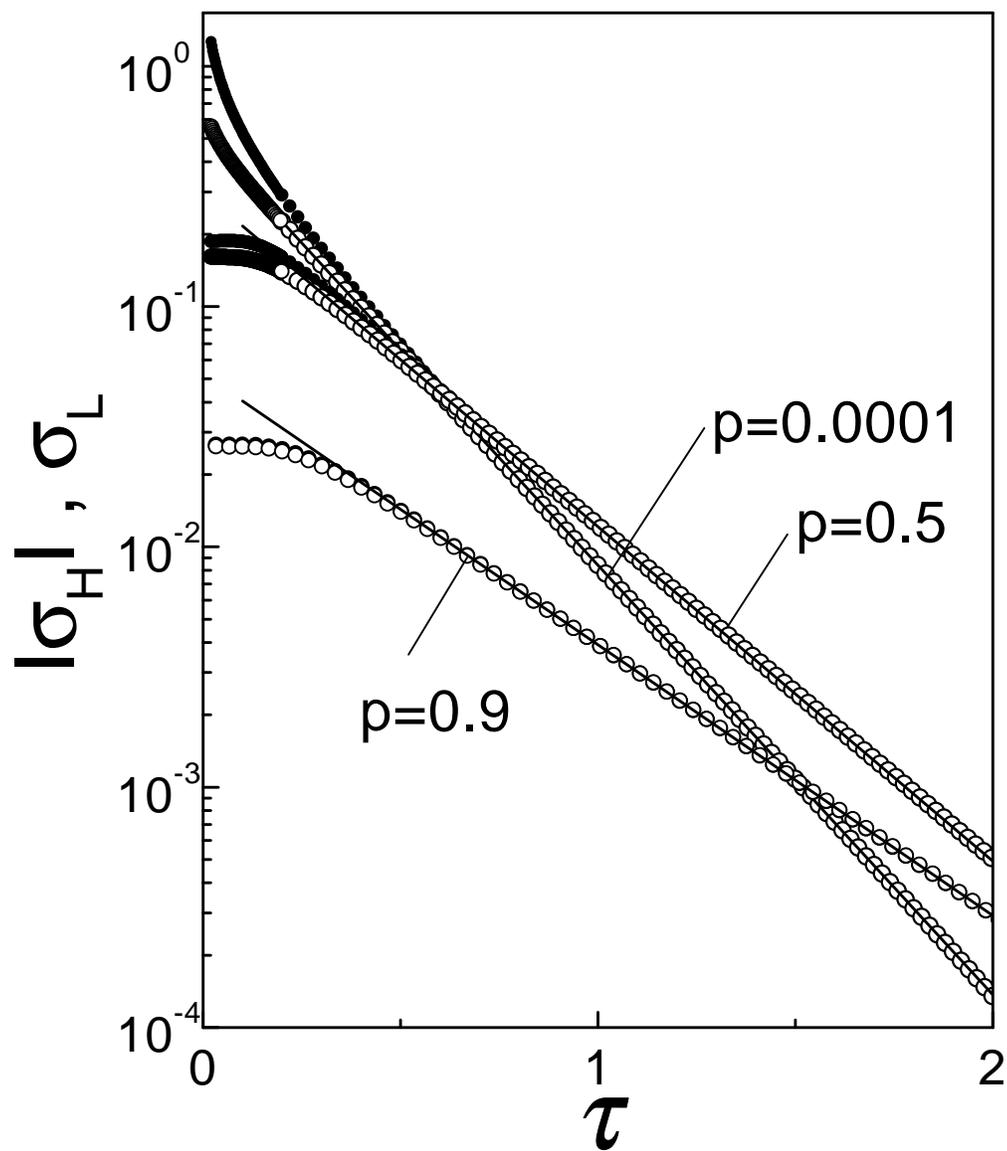
\end{document}